\documentclass[11pt]{article}

\title{Computer Science Challenges in Quantum Computing: \\ Early Fault-Tolerance and Beyond}

\author{
     Jens Palsberg \\ UCLA \\ palsberg@ucla.edu 
\and Jason Cong \\ UCLA \\ cong@cs.ucla.edu 
\and Yufei Ding \\ UCSD \\ yufeiding@ucsd.edu \\
\and Bill Fefferman \\ University of Chicago \\ wjf@uchicago.edu
\and Moinuddin Qureshi \\ Georgia Tech \\ moin@gatech.edu 
\and Gokul Subramanian Ravi \\ University of Michigan \\ gsravi@umich.edu \\
\and Kaitlin N. Smith \\ Northwestern University \\ kns@northwestern.edu 
\and Hanrui Wang \\ UCLA \\ wang@cs.ucla.edu 
\and Xiaodi Wu \\ U Maryland \\ xiaodiwu@umd.edu \\
\and Henry Yuen \\ Columbia University \\ henry.yuen@columbia.edu 
}
\date{January 25, 2026}

\usepackage{graphicx,url,hyperref}
\usepackage[dvipsnames,svgnames]{xcolor}
\usepackage[most]{tcolorbox}

\begin{document}

\maketitle

\newpage

\section*{Executive Summary}

Quantum computing is entering a period in which progress will be shaped as much by advances in computer science as by improvements in hardware. The central thesis of this report is that early fault-tolerant quantum computing shifts many of the primary bottlenecks from device physics alone to computer-science–driven system design, integration, and evaluation. While large-scale, fully fault-tolerant quantum computers remain a long-term objective, near- and medium-term systems will support early fault-tolerant computation with small numbers of logical qubits and tight constraints on error rates, connectivity, latency, and classical control. How effectively such systems can be used will depend on advances across algorithms, error correction, software, and architecture. This report identifies key research challenges for computer scientists and organizes them around these four areas, each centered on a fundamental question.

\paragraph{Algorithms, complexity, and cryptography.}
A central question is: {\em What problems admit quantum advantage?} Progress depends on identifying computational tasks for which quantum computers can outperform classical methods under realistic models and assumptions. This requires developing new algorithms, clarifying their limitations through complexity theory, and understanding how cryptographic or average-case assumptions can provide credible evidence for advantage when unconditional separations are out of reach.

\paragraph{Error correction and fault tolerance.}
A central question is: {\em Can quantum error correction be designed and deployed at scale?} Progress depends on moving beyond handcrafted constructions toward automated design and integration of codes, decoders, and architectures. This includes scalable decoding, principled resource trade-offs, and abstractions that support systematic co-design across hardware, software, and applications.

\paragraph{Software.}
A central question is: {\em Can software run efficiently and reliably across diverse hardware?} Progress depends on programming languages, compilers, and runtimes that support fault tolerance, heterogeneity, and hybrid execution without sacrificing correctness or performance. Advances in this area will determine how accessible, portable, and reliable quantum programming becomes as systems evolve.

\paragraph{Architecture.}
A central question is: {\em Can domain-specific machines enable earlier usefulness?} Progress depends on designing architectures that are co-designed with software and error correction and optimized for particular workload classes. Domain-specific designs may enable earlier practical impact while informing the development of more general-purpose systems.

\bigskip

Across these four areas, the report highlights themes that cut across the quantum computing stack. 
As quantum computations exceed classical simulability, {\bf verification and trust} become essential for relying on results in scientific and practical settings. 
At the same time, {\bf heterogeneity, co-design, and automation} reflect the growing diversity and complexity of quantum systems and the need for scalable integration across algorithms, software, and architecture. 
Complementing these research challenges, {\bf benchmarks and metrics} shape how progress is evaluated and compared under realistic constraints in the early fault-tolerant regime.
Together, these themes define a coherent research landscape for computer science in quantum computing.

\newpage

\begin{table}[t]
\centering
\small
\setlength{\tabcolsep}{6pt}
\renewcommand{\arraystretch}{1.2}
\begin{tabular}{p{2.95cm} p{3.77cm} p{3.8cm} p{4.2cm}}
\hline
\textbf{Area} &
\textbf{Framing question} &
\textbf{Early fault-tolerant bottlenecks} &
\textbf{Signals of progress} \\
\hline

\textbf{Algorithms, \mbox{complexity,~\&} cryptography} &
What problems admit quantum advantage? &
Evidence robust to dequantization; realistic models and assumptions &
Parameterized problem families with advantage evidence and clear baselines \\ \hline

\textbf{\mbox{Error correction} \& fault tolerance} &
Can QEC be designed and deployed at scale? &
Space--time overhead, decoding latency, classical control costs &
Automated code/decoder synthesis, and end-to-end, low-latency workflows \\ \hline

\textbf{Software} &
Can software run efficiently and reliably across diverse hardware? &
IRs and compilation for fault tolerance, heterogeneity, and dynamics &
Portable programs, transparent cost models, and verified transformations \\ \hline

\textbf{Architecture} &
Can domain-specific machines enable earlier usefulness? &
Mapping scarce logical qubits to structure and workloads &
Co-designed full-stack experiments showing application-level gains \\

\hline
\end{tabular}
\caption{Overview of the report.}
\label{tab:executive-map}
\end{table}

\newpage

\section{Introduction}

This report synthesizes the input of the community on the main quantum computing research challenges for computer scientists, with an emphasis on early fault-tolerant systems and areas where coordinated effort and shared evaluation practices can accelerate progress.
A key context for this report is recent breakthroughs in quantum error correction that have catapulted quantum computing research into a new era. Today, advances in computer science are increasingly central to the success of quantum computing. We have more exciting algorithms and more qubit technologies than ever, and we can now suppress hardware errors to some degree. In this diverse field, computer scientists can devise quantum algorithms and prove limits on what can be achieved; they can design and implement a software stack; and they can formally verify that algorithms and tools work correctly.
In early fault-tolerant regimes, space–time overhead, decoding latency, and classical control dominate; progress depends on automation, co-design, and trustworthy evaluation.
Such progress should be understood primarily as learning rather than as definitive demonstrations of advantage.
The remainder of the report motivates and explains the points summarized in Table~\ref{tab:executive-map}.

\paragraph{Scope.}
This report focuses on quantum computing; it does not address quantum sensing or quantum networking. It also does not address post-quantum cryptography, except where it informs quantum advantage and verification.

\paragraph{How to use this report.}
This report is intended to support the priority-setting and coordination in computer science research on quantum computing. It can be used to identify research directions where advances are most likely to influence the feasibility and usefulness of early fault-tolerant quantum systems, to balance short-term opportunities against longer-term foundational work, and to highlight shared challenges that benefit from coordinated effort. The questions, themes, and milestones outlined here are meant to be reference points rather than predictions and reflect the state of understanding at the time of writing; they may evolve as evidence accumulates and capabilities change. This report is intentionally technology- and institution-agnostic: it is meant to be useful to any organization planning research investments in quantum computing.
Several of the challenges outlined here may evolve faster or slower than anticipated; the value of the report lies in identifying structural bottlenecks rather than predicting timelines.

\paragraph{Workshop.}
We held a one-day \href{https://web.cs.ucla.edu/~palsberg/qcrc25.html}{workshop} with 90 attendees at the IEEE Quantum Week in Albuquerque on September 5, 2025. Our workshop had lightning talks in the morning, given by the presenters listed in the appendix, followed by breakout sessions and discussions in the afternoon. Each session was led by one of the co-authors of this report and had another co-author as the scribe.

\paragraph{Method.}
After the workshop, we assembled the scribe notes and polished them into a draft of this report. We asked for feedback on the draft report from the workshop speakers and additional community members.

\paragraph{Earlier reports.}
Over the years, the community has produced multiple forward-looking reports on quantum computing research challenges for computer scientists. We are particularly inspired by two reports from 2018 \cite{CRA18} and 2023 \cite{CRA23} that were sponsored by the Computing Community Consortium. Another report from 2017 \cite{QuantumSoftwareManifesto17} was written by members of the Quantum Software European Union Platform.

\paragraph{Acknowledgment.}
The authors thank both the workshop speakers and Frederic Chong, Alex Lombardi, and Charles Yuan for providing helpful comments on a draft of this report.
We are also grateful for the guidance of Margaret Martonosi and for financial support for the workshop from the United States \href{https://www.nsf.gov}{National Science Foundation} (\href{https://www.nsf.gov/awardsearch/showAward?AWD_ID=2526457&HistoricalAwards=false}{Award Number 2526457}) and the \href{https://www.ciqc.berkeley.edu}{Challenge Institute for Quantum Computation (CIQC)}.

\section{Trends and Projections}
\label{sec:trends-and-projections}

The landscape of quantum computing is evolving rapidly, driven by convergent trends in hardware capabilities, algorithmic discovery, and the fundamental shift toward error correction. In this section, we highlight key trajectories shaping the field, identify cross-cutting challenges that recur throughout the report, and offer a word of caution about interpreting early progress.

\subsection{Current Trajectories}

\paragraph{Advancing hardware technologies.}
The diversity of qubit technologies continues to expand, with viable platforms now including neutral-atom qubits, photonic qubits, semiconductor spin qubits, superconducting qubits, topological qubits, and trapped-ion qubits. Within each technology, we are witnessing a steady increase in hardware capability, characterized by growing numbers of physical qubits and continuous improvements in operation quality and gate fidelity. This hardware heterogeneity provides a rich physical substrate, but imposes significant demands on the system adaptability.

\paragraph{Increasing quantum applications.}
Parallel to hardware progress, every year sees new quantum algorithms and stronger evidence for an exponential advantage over classical computing in additional settings. These applications include integer factoring, physics simulation, and carefully constructed sampling and linear-algebra problems. Consequently, major efforts are underway to bridge theory and practice by estimating the concrete resources—gate counts and logical qubits—required to demonstrate practical advantage.  Notable examples include DARPA's \href{https://www.darpa.mil/research/programs/quantum-benchmarking-initiative}{QBI: Quantum Benchmarking Initiative}.

\paragraph{Paradigm shift from NISQ to fault tolerance.}
The field is undergoing a definitive paradigm shift from the Noisy Intermediate-Scale Quantum (NISQ) era to the era of early fault tolerance. There is now a broad consensus that Quantum Error Correction (QEC) is indispensable for scalable advantage. Industry and academia are racing to achieve full fault tolerance, with aggressive roadmaps targeting logical-level capabilities. For example, \href{https://www.quantinuum.com/press-releases/quantinuum-unveils-accelerated-roadmap-to-achieve-universal-fault-tolerant-quantum-computing-by-2030#}{Quantinuum's roadmap} projects fault-tolerant machines running millions of logical gates on hundreds of logical qubits by 2029; \href{https://www.ionq.com/roadmap}{IonQ's roadmap} targets systems with 80,000 logical qubits by 2030; and \href{https://www.ibm.com/roadmaps/quantum}{IBM's roadmap} envisions systems with 2,000 logical qubits capable of executing 1 billion gates by 2033+.
These roadmaps provide useful reference points, though their timelines and assumptions vary widely, and we interpret them as aspirational rather than predictive.

\paragraph{Emerging support from architecture and software.}
Despite these advances, a critical gap remains in the system stack. Previous work largely focused on the management of NISQ devices without error correction. As a result, the architectural abstractions and software tools required to support QEC-integrated systems are still underdeveloped. Developing a stack that can efficiently compile algorithms to diverse hardware platforms while managing programmer-selected error correction represents a frontier for computer science research.

\bigskip

The flurry of exciting developments has led to a broad set of quantum computing research challenges for computer scientists, which we outline in the following sections.  Progress on these challenges can help quantum computing to succeed sooner.  We will group the challenges under the headings of algorithms, complexity, and cryptography (Section~\ref{sec:alg-complexity-crypto}), error-correction and fault-tolerance (Section~\ref{sec:error-correction-and-fault-tolerance}), software (Section~\ref{sec:software}), and architecture (Section~\ref{sec:architecture}).  However, some challenges are cross-cutting and will be mentioned several times, particularly the following research challenges.

\paragraph{Verification and trust.}
As quantum computations move beyond the reach of classical simulation, verifying results and establishing trust in them becomes essential. This challenge spans cryptographic verification, formal reasoning about correctness, and system-level mechanisms that provide observability and well-characterized error bounds.
Different layers of the stack naturally require different verification techniques, ranging from cryptographic protocols and formal methods to statistical and system-level validation.

\paragraph{Heterogeneity, co-design, and automation.}
Quantum computing systems are likely to be heterogeneous in hardware technologies, error-correction schemes, and system organizations. Exploiting this heterogeneity may benefit from co-design across algorithms, software, and architecture, while automation—including artificial intelligence (AI) and machine-learning techniques—can help manage complexity in optimization, compilation, decoding, and system tuning.

\bigskip

Together, these cross-cutting themes provide a unifying perspective on the quantum computing research challenges for computer scientists.

\subsection{Towards Shared Benchmarks and Metrics}

Assessing progress in quantum computing requires benchmarks and metrics that reflect the realities of early fault-tolerant systems. As the field moves beyond NISQ devices, evaluation must capture how effectively complete systems execute meaningful algorithms under tight constraints on qubits, time, and classical control, in addition to raw hardware capability.

A useful benchmark suite should satisfy several canonical properties.
First, benchmarks should measure \emph{logical performance}, including logical error rates, error-correction overheads, latency, and classical processing costs, rather than focusing only on physical error metrics.
Second, benchmarks should be \emph{end-to-end}, exercising the full stack—from algorithms through compilation, error correction, and execution—so that trade-offs between layers are exposed clearly.
Third, benchmarks should be \emph{scalable and parameterized}, allowing problem size and system resources to grow together to track progress across hardware generations.
Fourth, benchmark suites should include a core of algorithms with credible evidence for exponential advantage under stated models and assumptions, alongside application-motivated workloads (e.g., chemistry, optimization, and machine learning) that stress realistic end-to-end behavior even when advantage is uncertain.
A balanced suite can therefore pair advantage-oriented kernels with system-oriented application workloads.
Including advantage-oriented kernels serves to anchor long-term goals, even when early fault-tolerant systems are primarily used to study scaling behavior, overheads, and failure modes rather than to demonstrate definitive advantage.

Benchmarks and metrics, therefore, serve as more than evaluation tools; they act as a coordinating mechanism for the field. They influence research priorities, enable comparison between platforms, and shape expectations about early fault-tolerant capabilities. For this reason, considerations about benchmarking recur throughout the report, appearing in the context of algorithms, error correction, software, and architecture. Clear, shared, and well-motivated benchmarks are essential for interpreting progress in early fault-tolerant quantum computing and for aligning effort across a diverse research ecosystem.

\subsection{A Word of Caution}

The transition to early fault-tolerant quantum computing brings both new opportunities and new risks. Although momentum in the field is substantial, not all research directions or evaluation practices are equally aligned with the capabilities and constraints of near- and medium-term systems. The following considerations highlight areas where a degree of caution can help ensure that effort and attention remain productively focused.

\paragraph{Overinterpreting early demonstrations.}
Early fault-tolerant experiments will necessarily operate at small scales and under tightly constrained conditions. Isolated demonstrations of logical qubits, error suppression, or limited algorithmic executions are important milestones, but they should not be taken as direct indicators of scalability or broad applicability. Care is needed to distinguish progress that generalizes from progress that depends on highly specific assumptions about noise models, layouts, or workloads.

\paragraph{Premature convergence on a single stack.}
The current diversity of hardware platforms, error-correction approaches, and software systems is a strength, rather than a weakness. Early consolidation around a single architecture, code family, programming model, or tool-chain risks narrowing the design space before the most important trade-offs are well understood. The maintenance of room for exploration and comparison remains valuable, particularly in the early fault-tolerant regime, where costs and constraints dominate performance.

\paragraph{Benchmarks that obscure key trade-offs.}
Simple or aggregate metrics can be appealing, but they can hide the distinctions that matter most for early fault-tolerant systems, such as space–time trade-offs, latency constraints, or classical control overheads. Benchmarks and merit figures should be interpreted carefully and paired with a qualitative analysis that makes assumptions explicit and highlights where results may or may extrapolate.

\paragraph{Underestimating systems complexity.}
As quantum computers incorporate error correction, classical control, runtime feedback, and heterogeneous components, the complexity of the system will grow rapidly. Progress in individual layers—algorithms, codes, compilers, or hardware—does not automatically translate into end-to-end capability. Sustained attention to integration, verification, and co-design is essential to avoid brittle systems that perform well in isolation but poorly in practice.

\paragraph{Expecting short-term clarity on quantum advantage.}
Determining which problems admit practical quantum advantage is likely to remain an incremental and iterative process. Early fault-tolerant machines may provide valuable data and insight without immediately resolving this question conclusively. Framing progress as learning—about costs, limits, and failure modes—can be more productive than expecting definitive demonstrations on short timescales.

\bigskip

Together, these considerations suggest prioritizing research that exposes scaling behavior, integration costs, and verification challenges over efforts that narrowly optimize for isolated demonstrations or idealized models. Recognizing the limits of early fault-tolerant systems can help align expectations, encourage robust evaluation practices, and support research choices that remain valuable as the field continues to evolve.

\section{Algorithms, Complexity, and Cryptography}
\label{sec:alg-complexity-crypto}

Quantum computer science is deeply rooted in the traditions and paradigms of theoretical computer science. After all, Shor's algorithm~\cite{Shor94} was inspired by Simon's algorithm~\cite{Simon96} which was inspired by finding query complexity evidence that quantum algorithms can be more powerful than classical algorithms. The original theoretical foundations of quantum advantage and the very concept of efficient quantum computation were formalized in Bernstein and Vazirani's seminal paper on quantum complexity theory~\cite{bernstein1997quantum}. 

Now, with the field making the jump from theory to practice, the study of quantum algorithms, complexity, and cryptography has developed a broader scope. As we will describe below, research in these areas will not just address fundamental theoretical and mathematical questions, but also build important connections with emerging areas of quantum computing such as verification, fault-tolerance, architecture, compilation, and more.

\subsection{Algorithms and Complexity}

The most important question in quantum algorithms and quantum complexity over the past thirty years has been the following: \emph{What problems admit quantum advantage?} 

Given the enormous interest and resources poured into quantum computing today, this question remains exceptionally important. Research in algorithmic foundations and theoretical computer science is critical for rigorously studying this question and must continue.
As quantum computing hardware becomes more reliable and readily available, there is an exciting opportunity for experimental and empirical methods to provide a deeper understanding of quantum algorithms and complexity. 

\paragraph{Beyond the black box model.} Much research on quantum advantage has been conducted in the \emph{black box} (also known as \emph{oracle}) model, where the input to a problem is accessed via black box queries. The most well-known quantum algorithms are all oracle algorithms (Grover search, Shor's algorithm, etc.). Strong separations between classical and quantum computation have been proven in the black box model (e.g., the Forrelation problem~\cite{aaronson2010bqp,raz2022oracle}). A recent breakthrough gave a separation relative to a random, rather than structured, oracle \cite{yamakawa2024verifiable}.  While oracle results can be often illuminating, they also have limitations and it would be desirable to study the nature of quantum speedups in ``white box'' models, where algorithms are allowed to analyze their input in more general ways; this would better match how classical and quantum algorithms operate in the real world. 

Recent work has begun to look at quantum advantage in the white box setting (e.g., decoded quantum interferometry~\cite{jordan2025optimization}, planted $k$-XOR~\cite{schmidhuber2025quartic}, guided local Hamiltonians~\cite{gharibian2022dequantizing,cade2023improved}, continuous optimization \cite{leng2025subexponentialquantumspeedupoptimization}, variants of Simon's problem \cite{ilango2025cryptography}, etc.). Although proving unconditional quantum advantage may be beyond the reach of current techniques in complexity theory, researchers should explore ways of giving strong evidence for quantum advantage in the white box model. For example, can we prove that the Decoded Quantum Interferometry problem admits a quantum advantage under a well-studied cryptographic assumption? Alternatively, can we prove that large classes of \emph{classical} algorithms (sum of squares, belief propagation, etc.) fail to match the performance of quantum algorithms for some problem?


\paragraph{Importing tools from information theory, statistical physics, and optimization.} Much of theoretical quantum computer science has been studied in the \emph{worst-case} setting, which means that algorithms, complexity, and cryptographic protocols are evaluated according to the worst-case \mbox{inputs/scenarios}. In many situations, this criterion may be too pessimistic. \emph{Average-case} analysis, on the other hand, can better model the behavior of algorithms in natural settings where input is drawn from some distribution. This paradigm is more effective in information theory, optimization, and AI.  

Additional research on quantum algorithms in the average-case setting is warranted, especially since this may better illustrate areas for quantum advantage. Tools and ideas from statistical physics, information theory, and optimization (such as belief propagation, message passing, replica method, and so on) have been powerful for studying average-case complexity~\cite{mezard2009information}; it would be interesting to see the effectiveness of these tools in the quantum setting. For example, such tools could potentially be brought to bear on the Decoded Quantum Interferometry (DQI) algorithm, which has recently received great interest~\cite{jordan2025optimization}; the algorithm is intrinsically connected to optimization, coding theory, and constraint satisfaction problems.  Another recent example is work on porting ideas from the so-called low-degree framework, a tool from the classical literature on average-case complexity of statistical inference tasks, into the average-case quantum learning setting \cite{chen2025informationcomputationgapsquantumlearning}.

\paragraph{Dequantization and limits of quantum advantage.} 
Absent unconditional complexity results or other forms of mathematical evidence, there is greater danger that a claim of quantum advantage may fall apart due to a cleverly designed classical algorithm. This is a recurring pattern in quantum algorithms research, where a quantum algorithm (initially thought to exhibit an advantage over classical computers) is \emph{dequantized} into a classical algorithm with comparable performance~\cite{gilyen2018quantum,tang2019quantum,DBLP:conf/stoc/ChiaGLLTW20,kothari2025no,Chaudhary25}.

This ``cat-and-mouse'' game between the quantum algorithm designers and the classical algorithm designers is an important dynamic that sharpens our understanding of where quantum advantage can exist. This yields a win-win scenario: either we dequantize a quantum algorithm (in which case we discover new, faster classical algorithms), or we have gained additional evidence for a quantum speedup (see, e.g., \cite{tang2019quantum,Fefferman-quantuminspired}).

We believe it is worthwhile to try to develop principled, theoretical frameworks for understanding when quantum algorithms can be dequantized, or maintain quantum advantage.


\paragraph{Quantum algorithms for quantum problems.} Much of quantum algorithms research has historically focused on classical input/output tasks: given a classical description of a problem instance, a quantum algorithm produces a classical solution.  Recently, there has been growing interest in \emph{intrinsically quantum} problems, where the inputs and outputs are quantum states, processes, or devices.  Examples include learning and testing quantum states and channels \cite{PRXQuantum.4.040337,ChenCutlerHuang25,10756089,10.1145/3717823.3718191,pmlr-v291-chen25c}, characterizing many-body systems, certifying quantum devices, and manipulating large entangled resources.  These questions demand new complexity-theoretic foundations: we need reductions and completeness notions for quantum tasks, robust measures of sample and query complexity when data is quantum, and models that capture tasks such as state and unitary synthesis (see~\cite{bostanci2023unitary,rosenthal2022interactive,metger2023stateqip,DBLP:conf/stoc/LombardiMW24,chia2024complexity} for initial explorations of such foundations).  Developing this theory will not only clarify which ``quantum-for-quantum'' tasks are achievable but will also connect to other areas including Hamiltonian complexity, learning theory, and quantum cryptography.


\paragraph{Algorithms and complexity under early fault-tolerant constraints.} In parallel, the emergence of early fault-tolerant quantum systems—operating below threshold but with limited logical qubits, depth, and connectivity—motivates theoretical models that more accurately capture practical constraints beyond the idealized fault-tolerant gate model.  Unlike the idealized, fully fault-tolerant gate model, these systems may support only restricted circuit depth, limited connectivity, components subject to residual noise after error correction, or particular families of native operations (such as analog Hamiltonian evolutions or specific multi-qubit gates~\cite{nadimpalli2024pauli}).  

We need complexity-theoretic frameworks
that can express these constraints and still ask sharp questions about
quantum advantage: for instance, which computational tasks can be
solved with early fault-tolerant quantum systems that have limited resources such as qubits or circuit depth but still remain hard for
classical algorithms (see, e.g., \cite{aaronson-linearoptical,fefferman-linearmodes,BFNV19,Bravyi-Gosset-Konig})?  Such models should inform both algorithm design and hardware development, guiding where to invest effort in state preparation, entangling operations, or error suppression. To complement these hardness results, it is important to develop
a better understanding of how well classical algorithms can take
advantage of realistic noise to simulate quantum experiments \cite{Fefferman-boson-algorithm,Dorit-RCS-algorithm,schuster2024polynomialtimeclassicalalgorithmnoisy, Fefferman-nonunital}, that is, residual noise after error correction. Finally, we would like to see a theory of early fault-tolerant models with meaningful lower bounds~\cite{aaronson-linearoptical,DBLP:conf/focs/BoulandFLL21, chen2023complexity,BoulandFefferman2, parham2025quantum} and separations that may inform the development and use of early fault-tolerant devices~\cite{DBLP:journals/iacr/FeffermanGSY25,foxman2025random}.


\paragraph{Co-design of classical and quantum algorithms.}
Some classical computations can benefit from quantum subroutines.  Recently, this has been demonstrated in an algorithm that does a classical diagonalization of a Hermitian matrix, taking advantage of quantum samples 
\cite{yu2025quantumcentricalgorithmsamplebasedkrylov}.  Another example is a recent biomarker-discovery algorithm that uses a quantum subroutine for feature selection \cite{shah2025quantumenabledbiomarkerdiscoveryoutlook}.  This kind of co-design is a promising direction for achieving an empirical quantum advantage.

\paragraph{Other quantum algorithmic paradigms.}
Finally, quantum algorithms are not limited to the standard circuit paradigm.  Alternative frameworks such as quantum walks, adiabatic and analog computation, continuous-variable and linear-optical models, distributed and communication-based quantum algorithms, and measurement-based computation all offer distinct ways to harness quantum mechanics (see, e.g., \cite{Childs-walk,Farhi-adiabatic,aaronson-linearoptical, leng2020qhd,Leng2025expandinghardware}).  Each paradigm comes with its own algorithmic techniques, complexity measures, and implementation trade-offs, and may align more naturally with particular hardware platforms or application domains.  We see opportunities for both deeper theory within each model and for comparative studies that relate these paradigms to each other and to the gate model.  A more systematic understanding of these diverse modes of quantum computation could reveal new algorithmic ideas, suggest architectures tailored to specific paradigms, and broaden the scope of problems where quantum advantage may plausibly be realized.

\subsection{Cryptography}


The last few years have been a renaissance for quantum cryptography, which can be roughly divided into two branches: \emph{post-quantum cryptography} (which concerns cryptosystems that classical computers can run, but are secure against quantum adversaries), and \emph{fully quantum cryptography} (where quantum information and quantum computation are used as part of a cryptosystem). 

Research in quantum cryptography has led to advances in quantum complexity theory, quantum algorithms, and quantum information. Thus, continued research in quantum cryptography will help us gain deeper insights into the nature of quantum information, quantum computation, and how we can make use of them for operational tasks.


\paragraph{Improved classical verification of quantum computation.} There has been tremendous progress on designing protocols to verify quantum computations with classical computers (or with classical computers augmented with weak quantum abilities). There are protocols with rigorous and formal security proofs, such as Mahadev's protocol~\cite{mahadev2022classical} and follow-up work like \cite{10.1007/978-3-031-07082-2_25}, but implementing these requires quantum computers with error rates significantly lower than what is achievable today. On the other hand, there are verification schemes that have been implemented on NISQ devices (such as the-so-called XEB test used in Google's quantum supremacy experiment in 2019~\cite{arute2019quantum}), but the security of these are heuristic and additionally require exorbitant classical resources. A significant and important research challenge is to try to close the gap between the two classes of protocols; the ideal goal would be to obtain a verification protocol with an efficient classical verifier that can be implemented on early fault-tolerant quantum hardware, and furthermore the security is proved under plausible cryptographic assumptions. 
A sub-challenge of full-fledged verification is achieving a more practical verifiable interactive quantum advantage. These are protocols where a prover can demonstrate quantum advantage under cryptographic assumptions such as Learning With Errors, at the cost of verification being interactive \cite{brakerski2021cryptographic,DBLP:conf/stoc/KalaiLV023,kahanamoku2022classically,alnawakhtha2024lattice}


\paragraph{Truly quantum primitives.} Quantum information enables new forms of cryptography that have no classical analog. Examples of this include quantum money/lightning/fire~\cite{aaronson2009quantum,zhandry2021quantum,bostanci2025general}, position verification~\cite{kent2011quantum,BuhrmanChandranFehrGellesGoyalOstrovskySchaffner14}, quantum copy-protection~\cite{aaronson2009quantum}, unclonable encryption~\cite{broadbent2020uncloneable}, one-shot signatures~\cite{amos2020one,shmueli2025one}, pseudorandom unitaries \cite{chen2024efficient,metger2024simple,ma2025construct}, and more. How can we construct these primitives and how can we analyze their security? Making progress on these questions will illuminate fundamental ways in which quantum information differs from classical information, as well as develop new mathematical techniques to study quantum information and quantum computation. Furthermore, finding ways to translate theoretical constructions of the aforementioned cryptographic primitives to practice will open up new avenues of utility for quantum technologies.

\paragraph{The complexity-theoretic foundations of quantum cryptography.} Just as classical cryptography and classical complexity theory are deeply intertwined (e.g., most classical cryptography requires assuming $\mathsf{P} \neq \mathsf{NP}$), recent work has begun to uncover deep connections between quantum cryptography and quantum complexity theory. However, much is unknown: for example, can the security of quantum cryptographic primitives be related to the hardness of computing a boolean function, or does some quantum cryptography rely on ``fully quantum'' computational assumptions? This is deeply related to the question of ``unitary synthesis'' of Aaronson and Kuperberg~\cite{aaronson2007quantum}. Another exciting development is the discovery of quantum cryptographic primitives whose security does not depend on the existence of one-way functions or $\mathsf{P}$ vs $\mathsf{NP}$~\cite{kretschmer2021quantum,DBLP:conf/crypto/AnanthQY22,morimae2022quantum,brakerski2023computational,kretschmer2023quantum,DBLP:conf/stoc/LombardiMW24,DBLP:conf/stoc/KhuranaT24,DBLP:journals/iacr/FeffermanGSY25,DBLP:conf/stoc/KhuranaT25}. Developing and exploring this theory will give us a deeper understanding of the possibilities and limits of quantum cryptography.  

\subsection{Milestones}

\begin{itemize}
\item
{\em Conceptual:}
Development of problem formulations and complexity-theoretic frameworks that clarify which computational tasks admit quantum advantage under realistic, white-box, and early fault-tolerant models. Development of richer methods for verifying quantum computations. Discovery of novel quantum cryptographic tasks.

\item
{\em Integration:}
Construction of models that connect algorithm design and complexity assumptions with hardware-relevant constraints such as limited logical qubits, circuit depth, noise, and native operations. Making verification protocols practically implementable on early fault-tolerant hardware.

\item
{\em Evaluation:}
Establishment of criteria and evidence for quantum advantage that remain robust to improved classical algorithms, including systematic understanding of dequantization and classical baselines. 
\end{itemize}


Signals of progress in this area include the emergence of problem families supported by multiple, independent lines of evidence for quantum advantage; clarity about the assumptions under which advantage is claimed; and sustained resistance to improved classical algorithms over time.





\section{Error Correction and Fault Tolerance}
\label{sec:error-correction-and-fault-tolerance}

Useful and fault-tolerant quantum computation will require quantum error correction (QEC). Until recently, QEC research was largely theoretical, as fault-tolerant systems seemed far from experimental realization. Much of the community effort has been focused on techniques such as noise-aware compilation and quantum error mitigation \cite{NiuSanialBronn24}. More recently, hardware advances across multiple qubit modalities have allowed logical qubits to be demonstrated~\cite{google2023suppressing, bluvstein2024logical} and error rates to be suppressed~\cite{google2025quantum}. Research focused on the surface code, the qLDPC codes, and the color code is becoming more prevalent. Still, large-scale demonstrations of fault-tolerant quantum computation remain elusive as the best approaches leading to system scaling have yet to be defined. 
In the early fault-tolerant regime, a central challenge for error correction and fault tolerance is: {\em Can quantum error correction be designed and deployed at scale?}


Many challenges must be overcome before QEC is practical. \textbf{The key challenge on the computer scientist's side is how to bridge theoretical QEC protocol design with heterogeneous hardware implementation in a resource-efficient, extensible manner.} Currently, there is a disconnect between theory and practice. To illustrate, space-efficient qLDPC codes are promising but challenging to implement due to their complex connectivity requirements \cite{wang2024coprimebivariatebicyclecodes}. Similarly, emerging heterogeneous systems involving multiple qubit modalities need strategic system design to fully exploit the strengths of each platform. Furthermore, as new components (e.g., decoders~\cite{higgott2025sparse,wu2023fusion,delfosse2021almost,Clique_Ravi2022,knapen2025pinballcryogenicpredecodersurface,das2022afs,vittal2023astrea,DBLP:conf/asplos/AlavisamaniVA0Q24,eraser,flagproxynetwork,wang26,maurya2025decoder,das2022lilliput}, magic state factories~\cite{gidney2024magic}) emerge, a future-proof and modular evaluation framework is indispensable to quickly testing new protocols.

The computer scientist has the opportunity to play a vital role in the development of the QEC stack. The transition from small-scale manual optimizations to million-qubit systems in which QEC implementations are efficiently synthesized based on the target application and architecture will require systematic design, automation, and mindful application of the principles of software engineering. 
%
%
%
Open-ended questions for computer scientists include the following:
\begin{itemize}
\item
\textbf{Architectural Interfaces and Co-design:} What abstractions and interfaces enable effective co-design across hardware, software, and applications? Specifically, how do we build modular frameworks that encapsulate hardware complexity to facilitate collaboration with physicists, while defining clear boundaries for system scalability and extensibility?
\item
\textbf{Automation and Scalability:} Where are the scalability limits of automated QEC design? Can classical computing, heuristics, and AI accelerate the discovery and synthesis of QEC implementations at million-qubit scales?
\item
\textbf{Verification and Benchmarking:} How should metrics balance fidelity, cost, and scalability? Which benchmarks best capture end-to-end performance, and how do we verify the correctness of execution on untrusted quantum devices?
\end{itemize}
In this section, we organize the key challenges for QEC into themes for future research efforts. These themes outline priority topics for computer science research and clarify points of interaction with adjacent research communities, both quantum and classical, to make meaningful progress.

\subsection{Architectural Foundations: Heterogeneity, Modularity, and Abstractions}

\paragraph{Hardware Diversity and Heterogeneity.}

Every existing qubit platform has unique strengths, and no single quantum technology will likely dominate all aspects of quantum computing. Computing systems comprise many components with unique roles, such as those used for processing, communication, and storage. With this in mind, future quantum systems may rely on heterogeneous architectures, where different qubit modalities play specialized roles. In classical systems, registers, caches, and memory hierarchies are fundamental and optimized for distinct purposes. Another similar analogy is the development of specialized components for commonly seen subroutines, like those for arithmetic building blocks. A similar approach may be necessary for quantum computing, with different QEC codes or hardware modalities tailored for specific tasks with different requirements for fidelity and latency. Such designs would leverage the strengths of multiple technologies, which may require a variety of different QEC codes. This raises many questions in terms of resource allocation, module sizing, conversion between qubit types and code types, and optimizations that provide context about coherence times and supported native gatesets, among other important properties that must be considered \cite{das2023imitation}. In addition, transient noise and system drift could add additional variation in quantum systems that evolve over time. The development of context-aware compilers developed specifically for heterogeneous systems may help bridge the wide gaps that exist between abstract programs, assorted QEC codes working concurrently, and heterogeneous hardware. Further, if flexible targets are possible for logical error rates, overheads can be kept manageable with an informed compiler, all while maintaining scalability. 

\paragraph{Distributed and Modular Systems.}

Constraints associated with power, cooling, and I/O will likely create ceilings with respect to scaling via a monolithic approach. However, modular design, where the subsystems are connected with quantum and classical interconnects, offers a potential solution to these scaling bottlenecks~\cite{smith2022scaling,laracuente2025modeling,tannu2017taming,zhang2025switchqnet, zhang2024mech}. When taking a modular approach, one can consider two pictures: either a local network of quantum processors or a large distributed system that spans distances that cover many kilometers. When considering a distributed approach for quantum computation, the additional degrees of freedom of the physical system create a substantial feature space when optimizing QEC. First, the question arises about how the system should be organized in terms of the logical qubit capacity of each included module, how codes can accommodate defects and device variation~\cite{lin2024codesign}, and how to estimate resource requirements. Additionally, we must answer questions regarding where magic state and entanglement factories are located and what their respective generation rates must be for target applications. Models that guide logical program scheduling will be paramount to ensure that needed resources are supplied just in time. Finally, a distributed quantum system is fundamentally hybrid as both quantum and classical elements are required. This raises the question: what are the best ways for classical processors to be co-located with quantum processors, and how can memory be shared either among clusters or globally? Although distributed computing principles from the classical domain can provide some assistance in addressing these challenges in distributed quantum computing, ultimately new frameworks for QEC will be needed to utilize modular architectures effectively.

\paragraph{Collaboration with Physicists, Device Engineers, and Industry.} 

The contributions of the computer scientist to QEC will be tightly coupled with the insights of hardware experts, making co-design essential. Decisions about the best QEC approach must be informed by realistic noise models, communication graphs of systems, and fabrication constraints, along with algorithms. As an example, each qubit technology might have an optimal set of native gates and unique noise channels that might need to be considered when designing a QEC framework~\cite{lin2022let}. Additionally, it must be understood how tolerant the most promising QEC codes are to defects and variations that are unavoidable in real hardware~\cite{palmer2025boundaries}. Although it may be impossible to consider all the parameters of the device when designing a QEC, the most important ones must be selected to inform system-level design while maintaining tractability of the optimization.

Maintaining open lines of communication between academic and industrial research groups is essential, as these sectors bring complementary strengths.  For example, industry brings resources like funding and more rapid advances in hardware that is often the result of decades of the refinement of classical processor fabrication techniques, like in the case of superconducting and silicon-based qubits. Academic groups, on the other hand, can provide novel testbeds that explore the utility of more nascent technology, such as exotic qubits or hardware that natively realizes qudit encodings. These types of effort are often deemed “high-risk, high-reward” and are not prioritized on industry roadmaps because they may not have a high probability of predictable long-term progress. In efforts focused on improving QEC, both sectors can mutually benefit if research efforts enhance the work of the other, rather than staying in silo, or worse, in competition.  

\paragraph{Development of Abstractions and New QEC Codes.}

Choosing the right abstraction during computation helps one to reason about program semantics without being overly concerned with unnecessary details, such as those associated with the underlying hardware. It may be of advantage to manage QEC with its own unique abstractions rather than having it be an opaque layer buried in the stack. Modular abstractions for codes, decoders, and runtimes with well-defined interfaces would allow systems to scale systematically, similar to how modular software and hardware allowed classical computing to advance. The additional benefit of modularity would be the ability to easily exchange QEC components during the prototype stages so that performance can be evaluated with a target application in mind. Additionally, separating the QEC abstraction(s) could allow quantum programs to be mapped in two stages: algorithm descriptions to logical qubits and logical qubits to physical operations. Discovering the right interfaces between the very highest, middle QEC, and lowest layers of the quantum computing stack could benefit from valuable insight from the classical computing community.

While we are developing modular abstractions for QEC and corresponding techniques for interfacing, we must also continue the search for new QEC codes. A promising avenue for developing improved codes that are more flexible would be encouraging co-design efforts between computer scientists, mathematicians, physicists, device engineers, and domain experts that are reaching the upper classical limits with the best-known approaches. Future systems may require multiple codes running side by side, especially within the heterogeneous quantum computing model. Additionally, some codes may better serve specific logical qubit roles, such as those acting on either the memory or the compute plane. Collaboration between the theoretical and applied communities will be critical in developing codes that are robust and can work seamlessly together, without facing severe decoding bottlenecks.

\subsection{Design Automation and System Integration}


If QEC is to scale and if its design is accessible to domain experts interested in running their specialized applications on quantum hardware, automation is essential to optimize programs for the many complex parameters associated with applications and physical systems. Quantum algorithms inherently have their own needs in terms of communication, fidelity requirements, and resources, and each system can demonstrate a diverse (and sometimes variable) space of architectural constraints, especially if heterogeneous and modular hardware is leveraged. This complexity creates the demand for tools and compilers that can explore massive parameter spaces to optimize resource allocation and promote low-latency decoding. Computer scientists have a natural advantage here. Just as electric design automation (EDA) accelerated the growth of classical computing to the point that it being ubiquitous, quantum design automation (QDA) could make the number of choices in fault-tolerant quantum systems more manageable. The challenge is how to integrate QDA across the quantum software stack, from compilers and runtime environment to QEC codes and hardware design. If QDA development is prioritized, scalability could be an early design principle rather than an afterthought. In addition, QDA could enable faster prototyping of deployment pipelines for new QEC proposals. Additionally, QDA tools could assist in dynamic runtime environments, allowing QEC techniques to adapt when noise environments or workloads change. If QDA for QEC is to have impact, it must connect everything that participates in a quantum computing task, from high-level algorithms down to the physical hardware, in a closed-loop manner.

The development of streamlined QDA that is unified end-to-end for QEC requires addressing many specific challenges to complete the pipeline. Many layers of the fault-tolerant stack must be tightly coupled. At the logical level, QDA tools for QEC must automatically construct ancilla systems that enable interactions between qubits. This involves selecting and instantiating stabilizer measurement circuits, lattice surgery operations~\cite{wang2025tableau, kan2025sparo}, or teleportation resources that are best suited for target code and hardware, all while adhering to physical resource constraints that may exist. Given these logical level communication protocols and the blocks of logical computation that they unify, QDA must next synthesize and schedule physical-level circuits~\cite{yin2025qecc, wu2022synthesis}, mapping abstract operations onto hardware-constrained gate sequences while jointly optimizing latency, qubit usage, and error accumulation~\cite{fang2025caliqec, yin2024surf}. To support the rapid deployment of applications and to accelerate the development of new QEC protocols, QDA should also enable the automated construction of complete fault-tolerant experiments, translating code and selected decoder strategy specifications into executable workloads informed by accurate noise models and configuration files supplied from all the quantum and classical processors that are leveraged. Finally, once a system configuration is specified and full program translation is possible, QDA must provide automated performance evaluation across a standardized set of benchmarks. This closes the loop by quantifying logical error rates, computation overheads, and throughput and feeding these metrics back into the EDA tool to improve design space exploration. Addressing these challenges associated with system integration would transform QEC design from a largely manual process into a compiler-driven workflow capable of navigating the enormous parameter spaces of scalable fault-tolerant quantum systems.  

\subsection{Benchmarks, Metrics, and Verification}

Benchmarks and metrics play a central role in the interpretation of progress in quantum error correction, as discussed in Section \ref{sec:trends-and-projections}. For QEC, meaningful evaluation focuses on logical performance, including logical error rates, physical-qubit overheads, decoding latency and throughput, and classical control costs, evaluated on representative end-to-end workloads. Parameterized benchmarks drawn from areas such as chemistry, optimization, and machine learning can expose how code choice, decoder design, and architectural constraints interact in practice. Public reporting of benchmark results, together with referenced artifacts and explicit assumptions, supports transparency, reproducibility, and comparison between competing QEC approaches.

Error correction does not remove all errors, it decreases the number of errors.  For the purpose of proving the correctness of error correction, we need mathematical statements of what correctness means in terms of QEC preserving the semantics of more abstract quantum program descriptions. Investment in verified quantum logic synthesis could assist with this challenge. For such workloads, correctness must be established without relying on a full classical simulation. Formal methods from computer science—such as theorem provers, model checking, and translation validation—may provide inspiration for certifying correctness at different levels of abstraction. A systematic method to efficiently verify workloads that demonstrate, ideally useful, quantum advantage would not only accelerate adoption but create a foundation of trust in the results of a quantum computer.




\subsection{Milestones}

\begin{itemize}
\item
{\em Conceptual:}
Development of shared abstractions for error-correction codes, decoders, and fault-tolerant protocols that clarify trade-offs among fidelity, overhead, latency, and scalability.

\item
{\em Integration:}
Construction of automated workflows that connect code selection, decoding strategies, and architectural parameters to target applications and system constraints.

\item
{\em Evaluation:}
Establishment of end-to-end benchmarks and metrics that demonstrate logical error rates, decoding latency, and resource costs across full fault-tolerant executions.
\end{itemize}


Signals of progress in this area include reductions in manual tuning effort, demonstrable scalability of automated workflows, and end-to-end evaluations that expose trade-offs among fidelity, latency, and resource cost across different system designs.

\section{Software}
\label{sec:software}

In the early fault-tolerant regime, a central challenge for software research is: {\em Can software run efficiently and reliably across diverse hardware?} Another challenge is whether software can facilitate application discovery and development by providing intuitive abstractions and programming support.

\subsection{Programming Languages}

Programmers want to express their quantum algorithms in high-level programming languages with problem-oriented abstractions, rather than hardware-level abstractions.  Designing such languages requires moving  
beyond circuit-level thinking and towards richer abstractions.  One direction is to design general languages that support any kind of quantum algorithm, while another is to design domain-specific languages.  One kind of domain is application domains, including physics simulation (e.g.~\cite{peng-simuq-2022}), cryptography, optimization (e.g.~\cite{kushnir2024qhdoptsoftwarenonlinearoptimization,wilson22,DBLP:conf/iccad/LiuJSWG24}), linear algebra (e.g.~\cite{yuan2025cobble}), and others. Another kind of domain is computational paradigms of quantum computation, such as adiabatic, analog, circuit-based, continuous-variable, and measurement-based computation.  The goal is to have programming languages that support innovation, achieve wide adoption, and can run on any quantum computer.
Every language should have formal semantics that enables implementers and tool designers to reason about correctness.

High-level abstractions can be general and support ideas such as control flow~\cite{yuan2024qcm}, functions, recursion, data structures~\cite{yuan2022tower}, memory, arithmetic, entanglement~\cite{yuan2022twist}, synchronization, distributed programming, and hybrid quantum-classical computation.  Those abstractions can also be domain specific and support, for example, Hamiltonian simulation for specific interesting models in many-body physics, high-energy physics or quantum chemistry, and information processing tasks in optimization, communication, and signal processing.
We can go further and unify disparate language constructs, which can help drive the area towards a single language.  Such a language should serve the wide variety of needs of algorithm designers, compiler writers, and quantum educators. 

\subsection{Benchmarks and Metrics}

Benchmarks play a coordinating role in quantum software research by grounding languages, compilers, and runtimes in representative workloads, as emphasized in Section~\ref{sec:trends-and-projections}. In the early fault-tolerant regime, software benchmarks should reflect realistic algorithmic structure, error-correction overheads, and hybrid quantum–classical execution patterns. A shared suite of benchmark programs that implement algorithms believed to offer an exponential quantum advantage—spanning multiple domains of application and input sizes—can support comparison between languages and tool chains, as exemplified by efforts such as SupermarQ~\cite{tomesh2022supermarqscalablequantumbenchmark}. Providing benchmarks in multiple programming languages and exposing common intermediate representations enables evaluation of the full software stack, from compilation through fault-tolerant execution, while preserving portability across hardware platforms.

A simulator is the easiest way to implement a language and test programs.
Although simulators will not scale to early fault-tolerant workloads, they remain essential for language design, compiler validation, and hybrid debugging workflows.
Researchers have developed many approaches to simulation, including state-vector methods, tensor networks, Clifford simulators~\cite{10.1145/3567955.3567958}, near-Clifford techniques, matchgate simulators~\cite{hassman2025enhancingchemistryquantumcomputers}, and Pauli propagation~\cite{rudolph2025paulipropagationcomputationalframework}.  As quantum programming languages continue to evolve, the field will continue to need new simulators.  For example, we will need simulation support for distributed quantum computing and for entanglement distillation. Scaling classical simulators and integrating them into hybrid workflows will be vital for both research and practical deployment.
Extensive research is required to continue to reduce the cost of simulation, which can potentially be achieved via full stack domain-specific tailoring of simulation.

\subsection{Compilers, Intermediate Representations, and Optimization}
Quantum compiler research~\cite{HuangPalsberg24,tannu2018case,FNM,micro1,adapt,jigsaw,DBLP:conf/asplos/Tannu0AQ22,ayanzadeh2023frozenqubits,badrike23,DBLP:conf/dac/NiuHIJY24} faces several cross-cutting challenges. Compilation for fault tolerance, support for different hardware backends (superconducting, trapped ion, neutral atom, and others), and scaling to millions of qubits are all essential goals.  

The field would benefit from an open-source compiler that can be used by many different researchers.
The design of such a compiler can be inspired by LLVM and MLIR, which are highly successful in the context of compilers for classical computing.
Several examples of open full-stack infrastructures already exist, such as QIBO, Open Quantum Design, and QSCOUT, suggesting that collaborative infrastructure development is feasible.
Current industry compilers tend to be only partially open source.
An open-source compiler that has well-defined and extensible intermediate representations would be a good basis for collaborative research.   

The field needs intermediate representations that can capture features like mid-circuit measurement and error correction.  
Those intermediate representations should have formal semantics.
Error correction might be added in a separate phase of the compiler, in which case we need intermediate representations for programs both before and after error correction has been added.
Those intermediate representations may be quite different from quantum circuits.
Another consideration is that error correction is inherently dynamic, raising questions about the need for dynamic compilation \cite{NiuKokcuMitraSzaszHashimKalloorJongIancuYounis24}. 

Optimizations rely on structural assumptions, and optimization strategies that worked in the NISQ era (such as just-in-time compilation combined with error mitigation) may not carry over to early fault-tolerant settings, where compilation is dominated by logical layout, decoding latency, and resource scheduling rather than noise mitigation. Offline compilation with strong average-case performance, layered with application-specific optimizations, may be more appropriate. Because many compiler optimizations are computationally intensive, parallel and high-performance computing techniques will likely be required, even for problems of moderate scale.
Additionally, the field needs optimization methods at multiple abstraction levels: a) application-level, e.g. rewriting Hamiltonians~\cite{10.1145/3622781.3674178,DBLP:conf/iccad/LiuJSWG24}; b) program-level, e.g. rewriting structured control flow~\cite{yuan2024spire}; and c) circuit-level.
Optimization rules can be automatically generated and applied \cite{DBLP:conf/pldi/XuLPLPHMPAAJ22,PointingPadonJiaMaHirthPalsbergAiken24}.

Mapping logical abstractions to physical execution introduces another set of challenges. One perspective frames quantum error correction itself as a layer of abstraction, while others stress the need to separate academic and industrial requirements. Logical-level abstractions should remain generic, and compiler stacks should be layered like GCC or LLVM. Current community efforts, such as Qiskit, are promising but could have better low-level optimizations \cite{DBLP:conf/dac/LiangSCHLWQWHQS23,10402000}.

As compilation pipelines grow deeper and more automated—particularly in the presence of error correction and dynamic execution—ensuring that these transformations preserve program meaning becomes a central challenge, motivating the need for systematic analysis and verification.

\subsection{Analysis, Verification, and Synthesis}

Quantum computing needs formal methods even more than classical computing.  As quantum hardware scales, simulation-based debugging and correctness assurance becomes increasingly challenging.  Once quantum computers outperform classical computers on some tasks, simulation-based approaches become infeasible.  Instead, the quantum field will need techniques for the analysis and verification of a wide variety of components of the software stack, including quantum programs, quantum circuits, compilers, optimizers, error correction, measurement, classical feedback, dynamic lifting, and the control software that helps operate the quantum hardware. 

The field may try a broad suite of approaches to analysis and verification, including type systems, static analysis \cite{DBLP:conf/pldi/YuP21, Peng-PLDI22,hung19qrobust}, program logic such as Hoare logic \cite{ying-popl17}, model checking, SMT solving \cite{li_et_al:LIPIcs.ECOOP.2024.24}, proof assistants such as Lean and Rocq, and testing on quantum devices using assertions, statistical methods~\cite{li-oracle-oopsla2022}, and connections to error correction.  For example, high-level languages with control flow and higher-order functions can benefit from type systems, while languages with support for uncomputation and short-lived temporary variables can benefit from static analysis.  
One can even develop fully certified and correct-by-construction compilers (e.g.,~\cite{hietala2021verified}) and quantum software (e.g.,~\cite{peng-shor-2023}) to achieve high-assurance implementation of large-scale practical quantum applications. 

Fundamentally, the field needs clearer definitions of the correctness of quantum programs, including specifications for intermediate states. 
Some applications, such as physics simulation, lack ground truth and new strategies are needed to verify intermediate representations.
Compilers should enforce application constraints to avoid missing simple but critical checks. Verification also ties directly to trust: one approach is to rely on third-party systems, smaller devices, or classical debugging queues. Stakeholders from different backgrounds view verification differently: physicists, software engineers, and end users may all emphasize different aspects.

A compiler can be buggy, and verifying that translation and program transformation preserve behavior is crucial.
Verification of compilers may be done either through verification of the compiler implementation itself or through translation validation that verifies the equivalence of a single source-target pair.  In this context, approximate implementations of unitaries pose unique challenges. 

Compilers can play a role in resource estimation even before the field reaches full fault tolerance.  The idea is that a compiler can produce target programs that, while too large to run on a quantum device and too large to simulate, can give a strong sense of the resource needed for execution.
This can be aided by cost models and program analyses that relate high-level quantum algorithms and programs to concrete architectural costs such as T-gates~\cite{yuan2024spire} and error correction cycles~\cite{harrigan2024}.  Researchers can also prove optimal lower bounds on resource needs 
\cite{PalsbergYu24,HuangPalsberg25}.

Program synthesis is the idea of mapping a specification of a task to a program that implements that task.  
For quantum computing, program synthesis can take the form of mapping a description of a unitary to a circuit that implements that unitary, either accurately or approximately.  A major challenge is to scale the synthesis beyond a few qubits and a few gates (e.g.,~\cite{deng-popl-2024,bugstahler24,wilson21}).

\subsection{Co-Design of Software and Hardware}

Software must increasingly account for the realities of quantum hardware architecture. Exploring the intersections between hardware design and software abstractions can improve performance and reliability
\cite{huo2026anchor,ludmir2024parallax,huo2026nest,ludmir2025modeling}.
The goal may be to make the implementation of high-level algorithms more aware of error correction and architecture~\cite{yin2025flexion, stein2025hetec}.
The field needs tools that enable programmers and compiler writers to work with QROM, QRAM~\cite{yuan2022tower}, and stored-program architectures~\cite{yuan2024qcm}.
The field also needs tools to work with different types of qubit: storage, computation, ancilla, dirty, etc.
Finally, the field may introduce domain-specific languages to build implementations of error-correcting codes. This could greatly aid efforts to verify implementations of such codes and to integrate them into the software stack.

\subsection{Database, Operating, and Networking Systems}

As the number of qubits grows, we may be able to think of a collection of qubits as a database and use them as such.  This will require us to develop techniques for quantum databases, including database models, indexing techniques \cite{DBLP:conf/vldb/GruenwaldWCGG23,fi16120439}, query languages, query processing, query optimization, security and privacy techniques, and transaction management \cite{Groppe22,DBLP:journals/pvldb/CalikyilmazGGWPSAPG23,Winker23}. 

The maturation of quantum computing depends on a robust quantum operating system and a control stack capable of translating high-level logical instructions into precise, real-time physical pulses to control quantum devices. 
At the higher levels of this stack, the fundamental tasks of quantum operating systems include resource management, scheduling, multiprogramming, and isolation in the presence of crosstalk \cite{PhysRevLett.125.150504,wang2024qoncord}. 
At the lower levels of this stack, the control system must execute quantum instructions, error-correction, and feedback loops within extremely short latency windows to maintain qubit coherence. 
RISC-Q~\cite{liu2025riscqgeneratorrealtimequantum} is a recent development that facilitates agile research into a lower-level operating system by serving as a RISC-V-compatible system-on-chip generator that enables the seamless hardware-software co-design of real-time quantum controllers.

Software support for quantum networking \cite{wehner2018quantum} is another example of operating systems for distributed execution environments, in which the operating system abstracts remote quantum processors, memories, and entanglement resources, enabling transparent scheduling, synchronization, and fault management across physically separated nodes. The fundamental tasks of a quantum network system~\cite{delle2025operating} include the distribution and management of entanglement~\cite{shi2020concurrent,clayton2024efficientroutingquantumnetworks}, the coordination of quantum and classical communication~\cite{10.1145/3341302.3342070}, the allocation of resources, and the enforcement of the timing and fidelity constraints required for reliable distributed quantum computation.

Other questions to be explored include the role of distributed compilation, multi-task execution~\cite{hou2025treevqatreestructuredexecutionframework,das2019case,NiuSanial22,NiuSanial23}, entanglement management, cloud quantum resource management~\cite{ravi2022quantumcomputingcloudanalyzing,ravi2022adaptivejobresourcemanagement}, and compilation for multi-quantum-processor programming~\cite{jeng2025modular}.

\subsection{AI for Quantum Software}

AI and machine learning are increasingly seen as tools to accelerate quantum software development~\cite{DBLP:conf/hpca/WangDGLPCH22, DBLP:conf/dac/0002G0LCP022, DBLP:conf/dac/0002LG0P022,DBLP:conf/asplos/Tannu0AQ22,DBLP:conf/dac/LiangL0C0LRSLY024}. Large language models and databases can support tasks such as circuit decomposition, test generation, and verification. Reinforcement learning has already been applied in circuit synthesis, mapping, and routing, and ML-based decoders are an active area of research. In early fault-tolerant devices, learning-based approaches can model hardware noise beyond simple stochastic assumptions, enabling adaptive error mitigation strategies that tailor circuit execution to time-varying and device-specific error profiles. Data-driven techniques can automatically select mitigation protocols, optimize measurement allocation, and infer latent noise parameters from limited experimental data, reducing the overhead required to extract reliable results from unreliable quantum hardware.
At the logical level, AI offers new opportunities for designing and optimizing quantum error correction codes. Machine learning can assist in discovering code structures with favorable thresholds, decoding complexity, or hardware compatibility, especially in regimes where analytical constructions are limited. Neural decoders and reinforcement-learning-based decoding policies can adapt to correlated and non-Markovian noise, outperforming fixed decoding rules under realistic conditions. More broadly, AI may enable co-design across hardware, codes, and decoders, jointly optimizing physical layouts, stabilizer measurements, and decoding strategies to maximize end-to-end logical fidelity.

In quantum compilation, learning-based methods have demonstrated promise in circuit synthesis, qubit mapping, scheduling, and pulse-level optimization~\cite{leng-daqc-neurips22,DBLP:conf/dac/LiangL0C0LRSLY024,10402000}. By learning cost models directly from hardware feedback, AI-assisted compilers can outperform hand-tuned heuristics in minimizing depth, error accumulation, and execution latency. Reinforcement learning and hybrid symbolic learning approaches are particularly attractive for navigating the combinatorial search spaces inherent in compilation while retaining the ability to respect hardware constraints and correctness guarantees. In addition, 
differentiable quantum programming can be investigated \cite{zhu-pldi20, fang-unbounded-2023,leng-daqc-neurips22}, inspired by its transformative role in classical machine learning to enable data-driven neural-symbolic quantum applications.
Such techniques are likely to become increasingly important as quantum systems scale and architectural heterogeneity grows.

Despite these advances, the integration of AI into quantum software raises fundamental questions about reliability, interpretability, and trust. Since quantum programs are often executed in regimes where errors are subtle but catastrophic, purely black-box AI outputs may be insufficient. Symbolic methods, formal verification, and constraint-based reasoning can complement learning-based approaches, providing correctness guarantees and enabling auditable AI-assisted workflows. Looking beyond compilers and control, AI will also play a critical role in managing and extracting insight from large quantum-generated datasets, such as those arising in quantum chemistry, materials science, and drug discovery. Effective data curation, representation, and validation will be essential to translate raw quantum outputs into scientifically meaningful results.


\subsection{Milestones}

\begin{itemize}
\item
{\em Conceptual:}
Development of programming languages and intermediate representations with precise semantics that support quantum error correction, modern quantum algorithms, mid-circuit measurement, and hybrid quantum–classical execution.

\item
{\em Integration:}
Construction of compiler tool chains that connect high-level programs to fault-tolerant executions across diverse hardware platforms using extensible representations and transparent cost models.

\item
{\em Evaluation:}
Establishment of practical methods and robust benchmarks for analysis, verification, and resource estimation that remain effective beyond classical simulation while fitting real software workflows.
\end{itemize}

Signals of progress in this area include increased portability of programs across platforms, improved transparency of resource costs, and verification techniques that scale beyond classical simulation while remaining usable in practice.

\section{Architecture}
\label{sec:architecture}

In the early fault-tolerant regime, a central challenge for architecture research is: {\em Can domain-specific machines enable earlier usefulness?}

\subsection{Domain-specific machines for fault-tolerant quantum computing}  

In the near- and medium term, available logical-qubit budgets (and the physical-qubit overhead required to obtain them) may be insufficient to support fault-tolerant execution of a wide array of problems.  So, can we enable reliable execution for at least one high-profile marquee problem?   Controlled demonstrations that illuminate scaling behavior and trade-offs will help to maintain a high level of attention to quantum computing over time.

One idea is to target domain-specific workload classes that share a small set of dominant kernels. Examples include (i) Hamiltonian simulation / time evolution (T-heavy with structured scheduling), (ii) structured linear algebra (repeated block-encodings, QROM-heavy data movement), and (iii) iterative hybrid workloads (mid-circuit measurement and tight classical feedback) \cite{jiao2025mediqganquantuminspiredganhigh}. These classes differ in whether they are limited primarily by T-state throughput, logical memory and data movement, or measurement/control latency, and each suggests a different architectural “shape.”

Domain-specific machines may become possible if the architecture, compiler, and code are specifically tailored for the given problem, rather than trying to solve the general case of trying to reliably execute a wide array of problems.  This type of effort can rally support from research and show a path of viability and commercial usefulness so that there is continued support for scaling quantum computing research to larger-scale machines.  Even classical systems took such a domain-specific approach, where the earliest machines, such as Enigma, were designed solely for the goal of breaking cryptographic codes.  There are two concerns with this approach. First, the selected problem should be such that it is indeed a viable candidate for quantum speedup (as we have often seen, once a problem becomes a candidate for quantum speedup, it usually draws attention from theoreticians, and classical algorithms improve to a point where the quantum speedup gets evaporated).  Second, the insights and solutions obtained from this approach should still be helpful in paving the way for a broader class of solutions. 

The specialization avenues below may be ways to optimize for such bottlenecks as T-state factories, communication/mapping, and control latency—while keeping an escape path to more general workloads.
\begin{itemize}
\item
Different magic state factory design depending on the available parallelism in the program.
\item
Instead of homogeneous connectivity, have regions with different connectivity and map different phases of programs to different regions, depending on the best fit,
such as those customized for QAOA~\cite{qaoa} and QCNN~\cite{cong2019qcnn} as shown in~\cite{lin2022dsqao}
\item Tailor mapping and routing to application requirements and application circuit design to fault-tolerance-related limitations, as shown, for example, in works that tailor early fault-tolerant quantum systems for VQAs~\cite{10.1145/3695053.3731112} and QML~\cite{elivagar,DBLP:conf/iccad/DiBritaLHLP24,dibrita2025resq}.
\item
Fault-tolerant quantum computing is resource-intensive and needs a large number of qubits, so there are plenty of opportunities for co-design. For example, different types of codes for different phases. 
\end{itemize}
These domain-specific efforts should be viewed as exploratory exemplars rather than commitments to a single long-term architecture.

\subsection{Benchmarks, Metrics, and Tools}

Shared benchmarks and metrics are essential to relate architectural design choices to application-level outcomes, as introduced in Section \ref{sec:trends-and-projections}. In the early fault-tolerant regime, architectural evaluation benefits from separating space, time, and classical control costs rather than collapsing them into a single figure of merit. Parameterized and automated benchmarks can reveal trade-offs among connectivity, modularity, code choice, and scheduling policies, even at scales smaller than full applications. Common simulation and evaluation tools that support end-to-end workflows enable consistent comparison across architectural designs and help connect low-level mechanisms to system-level performance.

\paragraph{Benchmarks.}  
Benchmarks must be parameterizable and automated.  In the early fault-tolerant regine, we are unlikely to have enough logical qubits to solve a commercially viable problem.  So, it would be useful to tailor the benchmark suite so that we have a relatively simpler set of problems and then, as machines scale, the problem complexity would increase.  For example, for classical machines, the initial set of benchmarks included simple kernels and toy problems (e.g. Towers of Hanoi), and then later on, the focus moved to commercially interesting problems. We could take a similar approach to quantum machines. The key challenge here would be to converge on the type of algorithms/benchmarks that would benefit from the error rates available in the near- and medium term. It would be useful to think about the intermediate steps needed to get to the final large-scale marquee application. It is also important that a benchmark suite has different types of workload with varied characteristics. For example, looking only at the Quantum Fourier Transform may give a wrong notion that all quantum programs are serial in nature with very little parallelism available across different instructions. Moreover, it is valuable to construct benchmarks with known optimal solutions so that we can measure the progress made by the community. The SupermarQ project is a disciplined first step in this direction~\cite{tomesh2022supermarqscalablequantumbenchmark}.

\paragraph{Metrics.} As the size of quantum computers scales and the applications executed on them evolve, it is important that the figure-of-merit used to evaluate and compare these machines must also adapt.  
Good progress has been made on architecture with fixed connectivity, such as QUEKO~\cite{queko} and QUBIKOS~\cite{qubikos}.
More work is needed for architectures with flexible connectivity, such as the trapped ion or neutral atom array.
This is especially important as we move from the NISQ model of computing (where the prevalent metric was the probability of successful trials) to other metrics that are more relevant to early fault-tolerant machines. Current metrics such as “Qubit Cycles” may be flawed, as space is much more valuable than time for early fault-tolerant machines, which will have limited qubit counts. Perhaps having metrics that separate space and time would be more meaningful for initial versions of quantum machines.  Using metrics from classical machines, such as the Iron law for performance (which separates execution time into three components: instructions, instructions-per-cycle, and cycle-time) may be helpful in providing insights for fault-tolerant quantum machines as well, especially in identifying which of the three components are affected by the given design change. 

\paragraph{Tools.} Research in computer architecture is guided by evaluations, and having a standard simulation infrastructure used by the community can help foster innovation quickly and facilitate comparisons of different designs. As the technology for future quantum machines is still in the exploration stage, it is crucial that such tools facilitate technology-aware evaluations,
such as the ZAC compiler for the evaluation of various configurations of a neutral atom-based architecture~\cite{zac}.
Ideally, the tools provide end-to-end evaluations of workloads under fault-tolerant quantum computing, such as the impact of QEC code, or the mapping, or the connectivity on the application reliability and the execution time.  Given the emergence of new space-efficient codes (qLDPC), it may also make sense to redo the analysis of quantum resource estimation for critical workloads (such as Shor) to see if such codes fundamentally change our understanding of when/how these applications become feasible.

\subsection{Towards Scalable Architectures} 

In classical computing, different components have well-defined functionality, such as interconnects, caches, and memory, and each of these is made with technologies that are specific to the component. The “sea of qubits” model for fault-tolerant quantum computing, where the same error correction code (and distance) is used for all logical qubits, is inefficient and difficult to scale. Instead, in the disaggregated model, a large majority of qubits are made with dense codes (akin to memory), a relatively smaller one are in hot storage area (similar to caches) that is made with surface code that they can participate in computation, and an even smaller area (akin to register files) that are used for creating magic states.  To enable such a disaggregated model, we need a good way to switch between different types of codes.  

As this paradigm is relatively new, it is useful to think about the architecture and management of different islands in the disaggregated design, for example, what is the level of connectivity required for each part, which technologies make sense, and so on (for the computation zone and the storage zone).  Furthermore, how do we partition the work between different zones, and between the classical machines and the quantum counterpart.  

Current technology exploration often separates qubit fabrication from control-system design.  Ideally, there should be a joint exploration of the quantum processor and controller to see if such an integrated design would provide better performance or reliability (due to the reduction in noise)
\cite{DBLP:conf/date/AlamA25,2024JAP...135a4903A,10.1063/5.0170187,Alam23,10.1063/5.0060716}.

Distributed quantum computing (DQC) is a key technology that can enable the disaggregated model. To enable DQC, we need a way to transmit data between different technologies in real time, so we may need architectural support to enable such low-latency transfers.  There are open problems in enabling quantum networks capable of enabling such communication.  Even for DQC, we want different islands to have high utilization so that the qubits on each of the sub-components remain useful for the overall computation.

As quantum systems scale, there is also a need to put in features of telemetry that can help in understanding the operation and performance of quantum machines.  Such features can improve the transparency of quantum architecture for domain scientists to debug their codes and build interfaces to collect run-time data for feedback. Adding visualization and explanation capabilities may enable domain scientists to better understand runtime errors, particularly those specific to the application. It would also be helpful to build abstractions and universal translation layers to virtualize heterogeneous quantum devices, enabling them to work seamlessly across different platforms. 


\subsection{Milestones}

\begin{itemize}
\item
{\em Conceptual:}
Development of architectural models for early fault-tolerant quantum systems that clarify space, time, and classical control costs and trade-offs among heterogeneity, modularity, and scalability.

\item
{\em Integration:}
Construction of domain-specific architectural designs that connect hardware organization, error correction, and compilation for targeted workload classes.

\item
{\em Evaluation:}
Establishment of shared evaluation frameworks and benchmark suites that demonstrate architectural trade-offs through reproducible, application-level comparisons.
\end{itemize}


Signals of progress in this area include architectural studies that connect design choices to application-level outcomes, reproducible comparisons using shared benchmarks, and designs that demonstrate clear benefits for targeted workload classes.

\section{Conclusion}

Quantum computing is entering a phase in which limited but genuine fault tolerance is becoming feasible. As this shift occurs, the central challenge of the field is changing. Progress increasingly depends on how effectively scarce logical qubits, time, and classical control can be used, rather than on device scale alone. In this regime, advances in computer science play a decisive role.

A central message of this report is that early fault-tolerant quantum computing should be viewed as a period of structured learning. Systems at this scale are unlikely to settle questions about broad quantum advantage, yet they can reveal the dominant costs, trade-offs, and failure modes that will shape future machines. Research that clarifies limits, enables systematic co-design, and supports trustworthy evaluation therefore has enduring value.

What would success look like over the next several years? Success would include reproducible executions of parameterized workloads, accompanied by transparent accounting of logical error rates, space--time overheads, and classical control costs. It would include software stacks that support fault tolerance, heterogeneity, and verification beyond classical simulation, enabling results to be interpreted and compared even when full validation is infeasible. It would also include shared benchmarks and metrics that make scaling behavior and system-level trade-offs visible across the stack.

At a broader level, success would be reflected in the emergence of stable abstractions and interfaces that make co-design routine rather than ad hoc, reducing reliance on manual tuning and bespoke assumptions. By treating early fault-tolerant computation as a guiding context, the research directions outlined in this report aim to support disciplined system-building and to lay a durable foundation for scalable and ultimately useful quantum computation.

\newpage

\section*{Presenters at the Workshop}

\begin{center}
\begin{tabular}{l}
\href{https://www.umut-acar.org}{Umut Acar} (Professor of Computer Science, Carnegie Mellon University) \\
\href{https://egr.vcu.edu/directory/tomasz.arodz}{Tomasz Arodz} (Professor of Computer Science, Virginia Commonwealth University) \\
\href{https://www.eecs.utk.edu/people/ahmedullah-aziz/}{Ahmedullah Aziz} (Professor of Computer Engineering, University of Tennessee, Knoxville) \\
\href{https://engineering.case.edu/about/school-directory/vipin-chaudhary}{Vipin Chaudhary} (Professor of Computer Science, Case Western Reserve University) \\
\href{https://www.sitanchen.com}{Sitan Chen} (Professor of Computer Science, Harvard University) \\
\href{https://sites.google.com/view/naihuichia}{Nai-Hui Chia} (Professor of Computer Science, Rice University) \\
\href{https://vast.cs.ucla.edu/people/faculty/jason-cong}{Jason Cong} (Professor of Computer Science, University of California, Los Angeles) \\
\href{https://www.ece.utexas.edu/people/faculty/poulami-das}{Poulami Das} (Professor of Electrical and Computer Engineering, University of Texas at Austin) \\
\href{https://cse.umn.edu/cs/caiwen-ding}{Caiwen Ding} (Professor of Computer Science, University of Minnesota) \\
\href{https://picassolab.squarespace.com/yufei}{Yufei Ding} (Professor of Computer Science, University of California, San Diego) \\
\href{https://www.billfefferman.com}{Bill Fefferman} (Professor of Computer Science, University of Chicago) \\
\href{https://cse.sc.edu/~pfu}{Frank Peng Fu} (Professor of Computer Science, University of South Carolina) \\
\href{https://www.ou.edu/coe/cs/people/faculty/le-gruenwald}{Le Gruenwald} (Professor of Computer Science, University of Oklahoma) \\
\href{https://www.gmu.edu/profiles/wjiang8}{Weiwen Jiang} (Professor of Computer Science, George Mason University) \\
\href{https://faculty.sites.iastate.edu/liyili2}{Liyi Li} (Professor of Computer Science, Iowa State University) \\
\href{https://sites.google.com/view/gushuli}{Gushu Li} (Professor of Computer Science, University of Pennsylvania) \\
\href{https://jerryzli.github.io}{Jerry Li} (Professor of Computer Science, University of Washington) \\
\href{https://www.qip-liu.com}{Qipeng Liu} (Professor of Computer Science, University of California, San Diego) \\
\href{https://www.depts.ttu.edu/cs/faculty/ying_liu/index.php}{Ying Liu} (Professor of Computer Science, Texas Tech University) \\
\href{https://www.unm.edu/~mmarvian}{Milad Marvian} (Professor of Electrical and Computer Engineering, University of New Mexico) \\
\href{https://arcb.csc.ncsu.edu/~mueller}{Frank Mueller} (Professor of Computer Science, North Carolina State University) \\
\href{https://bhabona.com/jobs/assistantship-phd-university-of-central-florida-orlando-florida-usa}{Siyuan Niu} (Professor of Electrical and Computer Engineering, University of Central Florida) \\
\href{https://web.cs.ucla.edu/~palsberg}{Jens Palsberg} (Professor of Computer Science, University of California, Los Angeles) \\
\href{https://www.ece.utexas.edu/people/faculty/david-z-pan}{David Pan} (Professor of Computer Science, University of Texas at Austin) \\
\href{https://www.tirthakpatel.com}{Tirthak Patel} (Professor of Computer Science, Rice University) \\
\href{https://pickspeng.github.io}{Yuxiang Peng} (Professor of Computer Science, Purdue University) \\
\href{https://moin.cc.gatech.edu}{Moinuddin Qureshi} (Professor of Computer Science, Georgia Institute of Technology) \\
\href{http://rand.cs.uchicago.edu}{Robert Rand} (Professor of Computer Science, University of Chicago) \\
\href{https://gsravi.engin.umich.edu}{Gokul Subramanian Ravi} (Professor of Computer Science, University of Michigan) \\
\href{https://engineering.nd.edu/faculty/yiyu-shi}{Yiyu Shi} (Professor of Computer Science, University of Notre Dame) \\
\href{https://www.mccormick.northwestern.edu/research-faculty/directory/profiles/smith-kaitlin.html}{Kaitlin Smith} (Professor of Computer Science, Northwestern University) \\
\href{https://caslab.io/jakub}{Jakub Szefer} (Professor of Computer Science, Northwestern University) \\
\href{https://xzt102.github.io}{Xulong Tang} (Professor of Computer Science, University of Pittsburgh) \\
\href{https://swamittannu.com}{Swamit Tannu} (Professor of Computer Science, University of Wisconsin, Madison) \\
\href{https://runzhoutao.github.io}{Runzhou Tao} (Professor of Computer Science, University of Maryland) \\
\href{https://www.chunhaowang.com}{Chunhao Wang} (Professor of Computer Science, University of California, Berkeley) \\
\href{https://people.eecs.berkeley.edu/~jswright}{John Wright} (Professor of Computer Science, University of California, Berkeley) \\
\href{https://www.cs.umd.edu/~xwu}{Xiaodi Wu} (Professor of Computer Science, University of Maryland) \\
\href{https://www.gmu.edu/profiles/lyang}{Lei Yang} (Professor of Computer Science, George Mason University) \\
\href{https://www.henryyuen.net}{Henry Yuen} (Professor of Computer Science, Columbia University) \\
\end{tabular}
\end{center}

\newpage

\section*{Glossary}

\paragraph{Dequantization.}
The discovery of a classical algorithm that matches or closely approximates the performance of a previously proposed quantum algorithm, calling into question the claimed quantum advantage.

\paragraph{Domain-specific quantum architecture.}
A quantum computer design that is tailored to run a particular class of algorithms or applications efficiently, rather than aiming to support all possible workloads equally well.

\paragraph{Early fault-tolerant computation.}
A stage of quantum computing in which error correction is used and errors are reduced below critical thresholds, but systems support only a small number of logical qubits and limited computation, with strong constraints on resources such as qubits, time, and classical control.

\paragraph{Heterogeneity in quantum systems.}
The use of different types of qubit, error-correction methods, or system components within a single quantum computing system, often to take advantage of their differing strengths.

\paragraph{Hybrid quantum--classical execution.}
A way of running programs in which quantum and classical computers work together during execution, for example, by using classical computation to guide or respond to measurements made on a quantum system.

\paragraph{Logical qubit.}
A qubit that is protected from errors using quantum error correction and implemented using many physical qubits, so that it behaves more reliably than any individual physical qubit.

\paragraph{Quantum advantage.}
A situation in which a quantum computer performs a task faster or more efficiently than the best known classical methods, under clearly stated assumptions and models.

\paragraph{Verification of quantum computation.}
Techniques for building confidence that a quantum computation is correct, including mathematical proofs, cryptographic checks, and statistical tests that can be applied when classical simulation is not feasible.

\newpage

\bibliographystyle{plain}
\bibliography{jens-references,quantum_papers,Ding,qureshi,wu}

@article{gidney2024magic,
  title={Magic state cultivation: growing T states as cheap as CNOT gates},
  author={Gidney, Craig and Shutty, Noah and Jones, Cody},
  journal={arXiv preprint arXiv:2409.17595},
  year={2024}
}

@article{wang2025tableau,
  title={Tableau-Based Framework for Efficient Logical Quantum Compilation},
  author={Wang, Meng and Liu, Chenxu and Garner, Sean and Stein, Samuel and Ding, Yufei and Nair, Prashant J and Li, Ang},
  journal={arXiv preprint arXiv:2509.02721},
  year={2025}
}

@inproceedings{fang2025caliqec,
  title={CaliQEC: In-situ Qubit Calibration for Surface Code Quantum Error Correction},
  author={Fang, Xiang and Yin, Keyi and Zhu, Yuchen and Ruan, Jixuan and Tullsen, Dean and Liang, Zhiding and Sornborger, Andrew and Li, Ang and Humble, Travis and Ding, Yufei and others},
  booktitle={Proceedings of the 52nd Annual International Symposium on Computer Architecture},
  pages={1402--1416},
  year={2025}
}

@inproceedings{zhang2025switchqnet,
  title={SwitchQNet: Optimizing Distributed Quantum Computing for Quantum Data Centers with Switch Networks},
  author={Zhang, Hezi and Xu, Yiran and Hu, Haotian and Yin, Keyi and Shapourian, Hassan and Zhao, Jiapeng and Kompella, Ramana Rao and Nejabati, Reza and Ding, Yufei},
  booktitle={Proceedings of the 52nd Annual International Symposium on Computer Architecture},
  pages={1449--1463},
  year={2025}
}

@article{kan2025sparo,
  title={SPARO: Surface-code Pauli-based Architectural Resource Optimization for Fault-tolerant Quantum Computing},
  author={Kan, Shuwen and Du, Zefan and Liu, Chenxu and Wang, Meng and Ding, Yufei and Li, Ang and Mao, Ying and Stein, Samuel},
  journal={arXiv preprint arXiv:2504.21854},
  year={2025}
}

@article{yin2025flexion,
  title={Flexion: Adaptive In-Situ Encoding for On-Demand QEC in Ion Trap Systems},
  author={Yin, Keyi and Fang, Xiang and Chen, Zhuo and Li, Ang and Hayes, David and Kaur, Eneet and Nejabati, Reza and Haeffner, Hartmut and Campbell, Wes and Hudson, Eric and others},
  journal={arXiv preprint arXiv:2504.16303},
  year={2025}
}

@inproceedings{stein2025hetec,
  title={HetEC: Architectures for Heterogeneous Quantum Error Correction Codes},
  author={Stein, Samuel and Xu, Shifan and Cross, Andrew W and Yoder, Theodore J and Javadi-Abhari, Ali and Liu, Chenxu and Liu, Kun and Zhou, Zeyuan and Guinn, Charlie and Ding, Yufei and others},
  booktitle={Proceedings of the 30th ACM International Conference on Architectural Support for Programming Languages and Operating Systems, Volume 2},
  pages={515--528},
  year={2025}
}

@inproceedings{yin2025qecc,
  title={QECC-Synth: A Layout Synthesizer for Quantum Error Correction Codes on Sparse Architectures},
  author={Yin, Keyi and Zhang, Hezi and Fang, Xiang and Shi, Yunong and Humble, Travis S and Li, Ang and Ding, Yufei},
  booktitle={Proceedings of the 30th ACM International Conference on Architectural Support for Programming Languages and Operating Systems, Volume 1},
  pages={876--890},
  year={2025}
}

@inproceedings{yin2024surf,
  title={Surf-Deformer: Mitigating dynamic defects on surface code via adaptive deformation},
  author={Yin, Keyi and Fang, Xiang and Humble, Travis S and Li, Ang and Shi, Yunong and Ding, Yufei},
  booktitle={2024 57th IEEE/ACM International Symposium on Microarchitecture (MICRO)},
  pages={750--764},
  year={2024},
  organization={IEEE}
}

@inproceedings{zhang2024mech,
  title={Mech: Multi-entry communication highway for superconducting quantum chiplets},
  author={Zhang, Hezi and Yin, Keyi and Wu, Anbang and Shapourian, Hassan and Shabani, Alireza and Ding, Yufei},
  booktitle={Proceedings of the 29th ACM International Conference on Architectural Support for Programming Languages and Operating Systems, Volume 2},
  pages={699--714},
  year={2024}
}

@inproceedings{wu2022synthesis,
  title={A synthesis framework for stitching surface code with superconducting quantum devices},
  author={Wu, Anbang and Li, Gushu and Zhang, Hezi and Guerreschi, Gian Giacomo and Ding, Yufei and Xie, Yuan},
  booktitle={Proceedings of the 49th Annual International Symposium on Computer Architecture},
  pages={337--350},
  year={2022}
}

@string{case = "International Workshop on CASE"}

@string{ecoop = "European Conference on Object-Oriented Programming"}

@string{ieee = "IEEE Press"}

@string{pldi="{ACM} {SIGPLAN} Conference on Programming Language Design and
                  Implementation"}

@string{Jan = "January"}

@string{Feb = "February"}

@string{May = "May"}

@string{Jun = "June"}

@string{Jul = "July"}

@string{Aug = "August"}

@string{Sep = "September"}

@string{Oct = "October"}

@string{Nov = "November"}

@string{Dec = "December"}

@string{PLDI    = "Proceedings of the ACM Conference on Programming Language
                     Design and Implementation"}

@string{SC      = "Proceedings of the ACM Conference on Supercomputing"}

@InProceedings{Shor94,
  title    = "Algorithms for Quantum Computation: 
              Discrete Logarithms and Factoring",
  author   = "Peter Shor",
  booktitle = "Proc. 35th Annu. Symp. on Foundations of Computer Science",
  publisher = "IEEE Press",
  pages    = "124--134",
  year     = 1994}

@InProceedings{Simon96,
  author    = "D. R. Simon",
  title     = "On the power of quantum computation", 
  booktitle = "Annual Symposium on Foundations of Computer Science",
  year      = 1996}

@Article{BuhrmanChandranFehrGellesGoyalOstrovskySchaffner14,
  author    = "Harry Buhrman and Nishanth Chandran and Serge Fehr and 
               Ran Gelles and Vipul Goyal and Rafail Ostrovsky and 
               Christian Schaffner",
  title     = "Position-Based Quantum Cryptography: 
               Impossibility and Constructions",
  journal   = "SIAM Journal of Computing",
  volume    = 43,
  number    = 1,
  pages     = "150--178",
  year      = 2014}

@InProceedings{HuangPalsberg24,
  author    = "Keli Huang and Jens Palsberg",
  title     = "\href{http://web.cs.ucla.edu/~palsberg/paper/pldi24.pdf}
                    {Compiling Conditional Quantum Gates 
                     without Using Helper Qubits}",
  booktitle = "Proceedings of PLDI'24, {ACM} {SIGPLAN} Conference on
               Programming Language Design and Implementation",
  month     = jun,
  year      = 2024}

@Article{HuangPalsberg25,
  author    = "Keli Huang and Jens Palsberg",
  title     = "Toffoli Requires Six Quantum Neighbor Gates",
  journal   = "ACM Transactions on Quantum Computing",
  note      = "Accepted; to appear",
  year      = 2026}

@Article{PalsbergYu24,
  author    = "Jens Palsberg and Nengkun Yu",
  title     = "Optimal Implementation of Quantum Gates with Two Controls",
  journal   = "Linear Algebra and Its Applications",
  volume    = 694,
  pages     = "206--261",
  doi       = "https://doi.org/10.1016/j.laa.2024.03.039",
  year      = 2024}

@article{hietala2021verified,
  title={A verified optimizer for quantum circuits},
  author={Hietala, Kesha and Rand, Robert and Hung, Shih-Han and Wu, Xiaodi and Hicks, Michael},
  journal={Proceedings of the ACM on Programming Languages},
  volume={5},
  number={POPL},
  pages={1--29},
  year={2021},
  publisher={ACM New York, NY, USA},
  doi = {https://doi.org/10.1145/3434318}
}

@Article{PointingPadonJiaMaHirthPalsbergAiken24,
  title = "Quanto: Optimizing Quantum Circuits with Automatic Generation of 
           Circuit Identities",
  author = "Jessica Pointing and Oded Padon and Zhihao Jia and Henry Ma and 
            Auguste Hirth and Jens Palsberg and Alex Aiken",
  journal = "Quantum Science and Technology",
  volume = 9,
  number = 4,
  year = 2024}

@TechReport{CRA18,
  author = "Margaret Martonosi and Martin Roetteler",
  title = "Next Steps in Quantum Computing:
Computer Science's Role",
  institution = "Computing Community Consortium",
  note = "\url{https://cra.org/ccc/wp-content/uploads/sites/2/2018/11/Next-Steps-in-Quantum-Computing.pdf}",
  year = 2018}

@TechReport{CRA23,
  author = "Kenneth Brown and Fred Chong and Kaitlin N. Smith and Thomas M. Conte and Austin Adams and Aniket Dalvi and Christopher Kang and Josh Viszlai",
  title = "5 Year Update to the Next Steps in Quantum Computing Report",
  institution = "Computing Community Consortium",
  note = "\url{https://cccblog.org/wp-content/uploads/2024/01/5-Year-Update-to-the-Next-Steps-in-Quantum-Computing.pdf}",
  year = 2023}

@TechReport{QuantumSoftwareManifesto17,
  author = "Andris Ambainis and Harry Buhrman and Elham Kashefi and
            Adrian Kent and Iordanis Kerenides and Frederik Kerling and
            Noah Linden and Ashley Montanaro and Floor van de Pavert and
            Thomas Strohm",
  title = "Quantum Software Manifesto",
  institution = "Quantum Software EU platform",
  year = 2017,
  note = "\url{https://www.eqsi.org/fileadmin/user_upload/lu_portal/eqsi.org/About/QSM_dd_18_nov.pdf}"}

@article{brakerski2021cryptographic,
  title={A cryptographic test of quantumness and certifiable randomness from a single quantum device},
  author={Brakerski, Zvika and Christiano, Paul and Mahadev, Urmila and Vazirani, Umesh and Vidick, Thomas},
  journal={Journal of the ACM (JACM)},
  volume={68},
  number={5},
  pages={1--47},
  year={2021},
  publisher={ACM New York, NY}
}

@article{kahanamoku2022classically,
  title={Classically verifiable quantum advantage from a computational Bell test},
  author={Kahanamoku-Meyer, Gregory D and Choi, Soonwon and Vazirani, Umesh V and Yao, Norman Y},
  journal={Nature Physics},
  volume={18},
  number={8},
  pages={918--924},
  year={2022},
  publisher={Nature Publishing Group UK London}
}

@article{alnawakhtha2024lattice,
  title={Lattice-based quantum advantage from rotated measurements},
  author={Alnawakhtha, Yusuf and Mantri, Atul and Miller, Carl A and Wang, Daochen},
  journal={Quantum},
  volume={8},
  pages={1399},
  year={2024},
  publisher={Verein zur F{\"o}rderung des Open Access Publizierens in den Quantenwissenschaften}
}

@article{yamakawa2024verifiable,
  title={Verifiable quantum advantage without structure},
  author={Yamakawa, Takashi and Zhandry, Mark},
  journal={Journal of the ACM},
  volume={71},
  number={3},
  pages={1--50},
  year={2024},
  publisher={ACM New York, NY}
}

@inproceedings{chen2024efficient,
  title={Efficient unitary designs from random sums and permutations},
  author={Chen, Chi-Fang and Docter, Jordan and Xu, Michelle and Bouland, Adam and Brand{\~a}o, Fernando GSL and Hayden, Patrick},
  booktitle={2024 IEEE 65th Annual Symposium on Foundations of Computer Science (FOCS)},
  pages={476--484},
  year={2024},
  organization={IEEE}
}

@inproceedings{metger2024simple,
  title={Simple constructions of linear-depth t-designs and pseudorandom unitaries},
  author={Metger, Tony and Poremba, Alexander and Sinha, Makrand and Yuen, Henry},
  booktitle={2024 IEEE 65th Annual Symposium on Foundations of Computer Science (FOCS)},
  pages={485--492},
  year={2024},
  organization={IEEE}
}

@inproceedings{ma2025construct,
  title={How to construct random unitaries},
  author={Ma, Fermi and Huang, Hsin-Yuan},
  booktitle={Proceedings of the 57th Annual ACM Symposium on Theory of Computing},
  pages={806--809},
  year={2025}
}

@inproceedings{ilango2025cryptography,
  title={Cryptography meets worst-case complexity: Optimal security and more from iO and worst-case assumptions},
  author={Ilango, Rahul and Lombardi, Alex},
  booktitle={2025 IEEE 66th Annual Symposium on Foundations of Computer Science (FOCS)},
  year={2025}
}

@inproceedings{10.1145/3695053.3731112,
author = {Dangwal, Siddharth and Vittal, Suhas and Seifert, Lennart Maximilian and Chong, Frederic T. and Ravi, Gokul Subramanian},
title = {Variational Quantum Algorithms in the era of Early Fault Tolerance},
year = {2025},
isbn = {9798400712616},
publisher = {Association for Computing Machinery},
address = {New York, NY, USA},
url = {https://doi.org/10.1145/3695053.3731112},
doi = {10.1145/3695053.3731112},
booktitle = {Proceedings of the 52nd Annual International Symposium on Computer Architecture},
pages = {1417–1431},
numpages = {15},
location = {
},
series = {ISCA '25}
}

@misc{tomesh2022supermarqscalablequantumbenchmark,
      title={SupermarQ: A Scalable Quantum Benchmark Suite}, 
      author={Teague Tomesh and Pranav Gokhale and Victory Omole and Gokul Subramanian Ravi and Kaitlin N. Smith and Joshua Viszlai and Xin-Chuan Wu and Nikos Hardavellas and Margaret R. Martonosi and Frederic T. Chong},
      year={2022},
      eprint={2202.11045},
      archivePrefix={arXiv},
      primaryClass={quant-ph},
      url={https://arxiv.org/abs/2202.11045}, 
}

@article{higgott2025sparse,
  title={Sparse blossom: correcting a million errors per core second with minimum-weight matching},
  author={Higgott, Oscar and Gidney, Craig},
  journal={Quantum},
  volume={9},
  pages={1600},
  year={2025},
  publisher={Verein zur F{\"o}rderung des Open Access Publizierens in den Quantenwissenschaften}
}

@inproceedings{10.1145/3622781.3674178,
author = {Seifert, Lennart Maximilian and Dangwal, Siddharth and Chong, Frederic T. and Ravi, Gokul Subramanian},
title = {Clapton: Clifford Assisted Problem Transformation for Error Mitigation in Variational Quantum Algorithms},
year = {2025},
isbn = {9798400703911},
publisher = {Association for Computing Machinery},
address = {New York, NY, USA},
url = {https://doi.org/10.1145/3622781.3674178},
doi = {10.1145/3622781.3674178},
booktitle = {Proceedings of the 29th ACM International Conference on Architectural Support for Programming Languages and Operating Systems, Volume 4},
pages = {47–62},
numpages = {16},
location = {Hilton La Jolla Torrey Pines, La Jolla, CA, USA},
series = {ASPLOS '24}
}

@misc{rudolph2025paulipropagationcomputationalframework,
      title={Pauli Propagation: A Computational Framework for Simulating Quantum Systems}, 
      author={Manuel S. Rudolph and Tyson Jones and Yanting Teng and Armando Angrisani and Zoë Holmes},
      year={2025},
      eprint={2505.21606},
      archivePrefix={arXiv},
      primaryClass={quant-ph},
      url={https://arxiv.org/abs/2505.21606}, 
}

@misc{ravi2022adaptivejobresourcemanagement,
      title={Adaptive job and resource management for the growing quantum cloud}, 
      author={Gokul Subramanian Ravi and Kaitlin N. Smith and Prakash Murali and Frederic T. Chong},
      year={2022},
      eprint={2203.13260},
      archivePrefix={arXiv},
      primaryClass={quant-ph},
      url={https://arxiv.org/abs/2203.13260}, 
}

@misc{ravi2022quantumcomputingcloudanalyzing,
      title={Quantum Computing in the Cloud: Analyzing job and machine characteristics}, 
      author={Gokul Subramanian Ravi and Kaitlin N. Smith and Pranav Gokhale and Frederic T. Chong},
      year={2022},
      eprint={2203.13121},
      archivePrefix={arXiv},
      primaryClass={quant-ph},
      url={https://arxiv.org/abs/2203.13121}, 
}

@misc{hou2025treevqatreestructuredexecutionframework,
      title={TreeVQA: A Tree-Structured Execution Framework for Shot Reduction in Variational Quantum Algorithms}, 
      author={Yuewen Hou and Dhanvi Bharadwaj and Gokul Subramanian Ravi},
      year={2025},
      eprint={2512.12068},
      archivePrefix={arXiv},
      primaryClass={quant-ph},
      url={https://arxiv.org/abs/2512.12068}, 
}

@misc{hassman2025enhancingchemistryquantumcomputers,
      title={Enhancing Chemistry on Quantum Computers with Fermionic Linear Optical Simulation}, 
      author={Zack Hassman and Oliver Reardon-Smith and Gokul Subramanian Ravi and Frederic T. Chong and Kevin J. Sung},
      year={2025},
      eprint={2511.12416},
      archivePrefix={arXiv},
      primaryClass={quant-ph},
      url={https://arxiv.org/abs/2511.12416}, 
}

@inproceedings{10.1145/3567955.3567958,
author = {Ravi, Gokul Subramanian and Gokhale, Pranav and Ding, Yi and Kirby, William and Smith, Kaitlin and Baker, Jonathan M. and Love, Peter J. and Hoffmann, Henry and Brown, Kenneth R. and Chong, Frederic T.},
title = {CAFQA: A Classical Simulation Bootstrap for Variational Quantum Algorithms},
year = {2022},
isbn = {9781450399159},
publisher = {Association for Computing Machinery},
address = {New York, NY, USA},
url = {https://doi.org/10.1145/3567955.3567958},
doi = {10.1145/3567955.3567958},
booktitle = {Proceedings of the 28th ACM International Conference on Architectural Support for Programming Languages and Operating Systems, Volume 1},
pages = {15–29},
numpages = {15},
location = {Vancouver, BC, Canada},
series = {ASPLOS 2023}
}

@inproceedings{wu2023fusion,
  title={Fusion blossom: Fast mwpm decoders for qec},
  author={Wu, Yue and Zhong, Lin},
  booktitle={2023 IEEE International Conference on Quantum Computing and Engineering (QCE)},
  volume={1},
  pages={928--938},
  year={2023},
  organization={IEEE}
}

@article{delfosse2021almost,
  title={Almost-linear time decoding algorithm for topological codes},
  author={Delfosse, Nicolas and Nickerson, Naomi H},
  journal={Quantum},
  volume={5},
  pages={595},
  year={2021},
  publisher={Verein zur F{\"o}rderung des Open Access Publizierens in den Quantenwissenschaften}
}

@INPROCEEDINGS{Clique_Ravi2022,
  doi = {10.48550/ARXIV.2208.08547},
  
  url = {https://arxiv.org/abs/2208.08547},
  
  author = {Ravi, Gokul Subramanian and Baker, Jonathan M. and Fayyazi, Arash and Lin, Sophia Fuhui and Javadi-Abhari, Ali and Pedram, Massoud and Chong, Frederic T.},
  
  title = {Better than worst-case decoding for quantum error correction},
  
  booktitle={ACM International Conference on Architectural Support for Programming Languages and Operating Systems (ASPLOS)},   
  year = {2023},
  
  copyright = {arXiv.org perpetual, non-exclusive license}
}

@misc{knapen2025pinballcryogenicpredecodersurface,
      title={Pinball: A Cryogenic Predecoder for Surface Code Decoding Under Circuit-Level Noise}, 
      author={Alexander Knapen and Guanchen Tao and Jacob Mack and Tomas Bruno and Mehdi Saligane and Dennis Sylvester and Qirui Zhang and Gokul Subramanian Ravi},
      year={2025},
      eprint={2512.09807},
      archivePrefix={arXiv},
      primaryClass={quant-ph},
      url={https://arxiv.org/abs/2512.09807}, 
}

@inproceedings{aaronson2009quantum,
  title={Quantum copy-protection and quantum money},
  author={Aaronson, Scott},
  booktitle={2009 24th Annual IEEE Conference on Computational Complexity},
  pages={229--242},
  year={2009},
  organization={IEEE}
}

@article{zhandry2021quantum,
  title={Quantum lightning never strikes the same state twice. or: quantum money from cryptographic assumptions},
  author={Zhandry, Mark},
  journal={Journal of Cryptology},
  volume={34},
  number={1},
  pages={6},
  year={2021},
  publisher={Springer}
}

@inproceedings{bostanci2025general,
  title={A General Quantum Duality for Representations of Groups with Applications to Quantum Money, Lightning, and Fire},
  author={Bostanci, John and Nehoran, Barak and Zhandry, Mark},
  booktitle={Proceedings of the 57th Annual ACM Symposium on Theory of Computing},
  pages={201--212},
  year={2025}
}

@article{kent2011quantum,
  title={Quantum tagging: {A}uthenticating location via quantum information and relativistic signaling constraints},
  author={Kent, Adrian and Munro, William J and Spiller, Timothy P},
  journal={Physical Review A—Atomic, Molecular, and Optical Physics},
  volume={84},
  number={1},
  pages={012326},
  year={2011},
  publisher={APS}
}

@inproceedings{broadbent2020uncloneable,
  title={Uncloneable {Q}uantum {E}ncryption via {O}racles},
  author={Broadbent, Anne and Lord, S{\'e}bastien},
  booktitle={15th Conference on the Theory of Quantum Computation, Communication and Cryptography},
  year={2020}
}

@inproceedings{amos2020one,
  title={One-shot signatures and applications to hybrid quantum/classical authentication},
  author={Amos, Ryan and Georgiou, Marios and Kiayias, Aggelos and Zhandry, Mark},
  booktitle={Proceedings of the 52nd Annual ACM SIGACT Symposium on Theory of Computing},
  pages={255--268},
  year={2020}
}

@inproceedings{shmueli2025one,
  title={On one-shot signatures, quantum vs. classical binding, and obfuscating permutations},
  author={Shmueli, Omri and Zhandry, Mark},
  booktitle={Annual International Cryptology Conference},
  pages={350--383},
  year={2025},
  organization={Springer}
}

@inproceedings{rosenthal2022interactive,
  title={Interactive {P}roofs for {S}ynthesizing {Q}uantum States and {U}nitaries},
  author={Rosenthal, Gregory and Yuen, Henry},
  booktitle={13th Innovations in Theoretical Computer Science Conference (ITCS 2022)},
  pages={112--1},
  year={2022},
  organization={Schloss Dagstuhl--Leibniz-Zentrum f{\"u}r Informatik}
}

@inproceedings{metger2023stateqip,
  title={{stateQIP} = {statePSPACE}},
  author={Metger, Tony and Yuen, Henry},
  booktitle={2023 IEEE 64th Annual Symposium on Foundations of Computer Science (FOCS)},
  pages={1349--1356},
  year={2023},
  organization={IEEE}
}

@inproceedings{nadimpalli2024pauli,
  title={On the {P}auli spectrum of {QAC0}},
  author={Nadimpalli, Shivam and Parham, Natalie and Vasconcelos, Francisca and Yuen, Henry},
  booktitle={Proceedings of the 56th Annual ACM Symposium on Theory of Computing},
  pages={1498--1506},
  year={2024}
}

@article{parham2025quantum,
  title={Quantum circuit lower bounds in the magic hierarchy},
  author={Parham, Natalie},
  journal={arXiv preprint arXiv:2504.19966},
  year={2025}
}

@article{mahadev2022classical,
  title={Classical verification of quantum computations},
  author={Mahadev, Urmila},
  journal={SIAM Journal on Computing},
  volume={51},
  number={4},
  pages={1172--1229},
  year={2022},
  publisher={SIAM}
}

@article{chen2023complexity,
  title={The complexity of {NISQ}},
  author={Chen, Sitan and Cotler, Jordan and Huang, Hsin-Yuan and Li, Jerry},
  journal={Nature Communications},
  volume={14},
  number={1},
  pages={6001},
  year={2023},
  publisher={Nature Publishing Group UK London}
}

@article{foxman2025random,
  title={Random unitaries in constant (quantum) time},
  author={Foxman, Ben and Parham, Natalie and Vasconcelos, Francisca and Yuen, Henry},
  journal={arXiv preprint arXiv:2508.11487},
  year={2025}
}

@article{chia2024complexity,
  title={Complexity {T}heory for {Q}uantum {P}romise {P}roblems},
  author={Chia, Nai-Hui and Chung, Kai-Min and Huang, Tzu-Hsiang and Shih, Jhih-Wei},
  journal={arXiv preprint arXiv:2411.03716},
  year={2024}
}

@article{DBLP:journals/iacr/FeffermanGSY25,
  author       = {Bill Fefferman and
                  Soumik Ghosh and
                  Makrand Sinha and
                  Henry Yuen},
  title        = {The Hardness of Learning Quantum Circuits and its Cryptographic Applications},
  journal      = {{IACR} Cryptol. ePrint Arch.},
  pages        = {718},
  year         = {2025},
  url          = {https://eprint.iacr.org/2025/718},
  bibsource    = {dblp computer science bibliography, https://dblp.org}
}

@article{DBLP:journals/pvldb/CalikyilmazGGWPSAPG23,
  author       = {Umut {\c{C}}alikyilmaz and
                  Sven Groppe and
                  Jinghua Groppe and
                  Tobias Winker and
                  Stefan Prestel and
                  Farida Shagieva and
                  Daanish Arya and
                  Florian Preis and
                  Le Gruenwald},
  title        = {Opportunities for Quantum Acceleration of Databases: Optimization
                  of Queries and Transaction Schedules},
  journal      = {Proc. {VLDB} Endow.},
  volume       = {16},
  number       = {9},
  pages        = {2344--2353},
  year         = {2023},
  url          = {https://www.vldb.org/pvldb/vol16/p2344-calikyilmaz.pdf},
  doi          = {10.14778/3598581.3598603},
  bibsource    = {dblp computer science bibliography, https://dblp.org}
}

@inproceedings{DBLP:conf/asplos/AlavisamaniVA0Q24,
  author       = {Narges Alavisamani and
                  Suhas Vittal and
                  Ramin Ayanzadeh and
                  Poulami Das and
                  Moinuddin K. Qureshi},
  editor       = {Rajiv Gupta and
                  Nael B. Abu{-}Ghazaleh and
                  Madan Musuvathi and
                  Dan Tsafrir},
  title        = {Promatch: Extending the Reach of Real-Time Quantum Error Correction
                  with Adaptive Predecoding},
  booktitle    = {Proceedings of the 29th {ACM} International Conference on Architectural
                  Support for Programming Languages and Operating Systems, Volume 3,
                  {ASPLOS} 2024, La Jolla, CA, USA, 27 April 2024- 1 May 2024},
  pages        = {818--833},
  publisher    = {{ACM}},
  year         = {2024},
  url          = {https://doi.org/10.1145/3620666.3651339},
  doi          = {10.1145/3620666.3651339},
  bibsource    = {dblp computer science bibliography, https://dblp.org}
}

@inproceedings{morimae2022quantum,
  title={Quantum commitments and signatures without one-way functions},
  author={Morimae, Tomoyuki and Yamakawa, Takashi},
  booktitle={Annual International Cryptology Conference},
  pages={269--295},
  year={2022},
  organization={Springer}
}

@inproceedings{brakerski2023computational,
  title={On the {C}omputational {H}ardness {N}eeded for {Q}uantum {C}ryptography},
  author={Brakerski, Zvika and Canetti, Ran and Qian, Luowen},
  booktitle={14th Innovations in Theoretical Computer Science Conference (ITCS 2023)},
  pages={24--1},
  year={2023},
  organization={Schloss Dagstuhl--Leibniz-Zentrum f{\"u}r Informatik}
}

@inproceedings{aaronson2007quantum,
  title={Quantum versus classical proofs and advice},
  author={Aaronson, Scott and Kuperberg, Greg},
  booktitle={Twenty-Second Annual IEEE Conference on Computational Complexity (CCC'07)},
  pages={115--128},
  year={2007},
  organization={IEEE}
}

@inproceedings{kretschmer2021quantum,
  title={Quantum {P}seudorandomness and {C}lassical {C}omplexity},
  author={Kretschmer, William},
  booktitle={16th Conference on the Theory of Quantum Computation, Communication and Cryptography (TQC 2021)},
  pages={2--1},
  year={2021},
  organization={Schloss Dagstuhl--Leibniz-Zentrum f{\"u}r Informatik}
}

@inproceedings{kretschmer2023quantum,
  title={Quantum cryptography in algorithmica},
  author={Kretschmer, William and Qian, Luowen and Sinha, Makrand and Tal, Avishay},
  booktitle={Proceedings of the 55th Annual ACM Symposium on Theory of Computing},
  pages={1589--1602},
  year={2023}
}

@article{arute2019quantum,
  title={Quantum supremacy using a programmable superconducting processor},
  author={Arute, Frank and Arya, Kunal and Babbush, Ryan and Bacon, Dave and Bardin, Joseph C and Barends, Rami and Biswas, Rupak and Boixo, Sergio and Brandao, Fernando GSL and Buell, David A and others},
  journal={Nature},
  volume={574},
  number={7779},
  pages={505--510},
  year={2019},
  publisher={Nature Publishing Group UK London}
}

@inproceedings{DBLP:conf/asplos/Tannu0AQ22,
  author       = {Swamit S. Tannu and
                  Poulami Das and
                  Ramin Ayanzadeh and
                  Moinuddin K. Qureshi},
  editor       = {Babak Falsafi and
                  Michael Ferdman and
                  Shan Lu and
                  Thomas F. Wenisch},
  title        = {{HAMMER:} boosting fidelity of noisy Quantum circuits by exploiting
                  Hamming behavior of erroneous outcomes},
  booktitle    = {{ASPLOS} '22: 27th {ACM} International Conference on Architectural
                  Support for Programming Languages and Operating Systems, Lausanne,
                  Switzerland, 28 February 2022 - 4 March 2022},
  pages        = {529--540},
  publisher    = {{ACM}},
  year         = {2022},
  url          = {https://doi.org/10.1145/3503222.3507703},
  doi          = {10.1145/3503222.3507703},
  bibsource    = {dblp computer science bibliography, https://dblp.org}
}

@inproceedings{DBLP:conf/crypto/AnanthQY22,
  author       = {Prabhanjan Ananth and
                  Luowen Qian and
                  Henry Yuen},
  editor       = {Yevgeniy Dodis and
                  Thomas Shrimpton},
  title        = {Cryptography from Pseudorandom Quantum States},
  booktitle    = {Advances in Cryptology - {CRYPTO} 2022 - 42nd Annual International
                  Cryptology Conference, {CRYPTO} 2022, Santa Barbara, CA, USA, August
                  15-18, 2022, Proceedings, Part {I}},
  series       = {Lecture Notes in Computer Science},
  volume       = {13507},
  pages        = {208--236},
  publisher    = {Springer},
  year         = {2022},
  url          = {https://doi.org/10.1007/978-3-031-15802-5\_8},
  doi          = {10.1007/978-3-031-15802-5\_8},
  bibsource    = {dblp computer science bibliography, https://dblp.org}
}

@inproceedings{DBLP:conf/dac/0002G0LCP022,
  author       = {Hanrui Wang and
                  Jiaqi Gu and
                  Yongshan Ding and
                  Zirui Li and
                  Frederic T. Chong and
                  David Z. Pan and
                  Song Han},
  editor       = {Rob Oshana},
  title        = {QuantumNAT: quantum noise-aware training with noise injection, quantization
                  and normalization},
  booktitle    = {{DAC} '22: 59th {ACM/IEEE} Design Automation Conference, San Francisco,
                  California, USA, July 10 - 14, 2022},
  pages        = {1--6},
  publisher    = {{ACM}},
  year         = {2022},
  url          = {https://doi.org/10.1145/3489517.3530400},
  doi          = {10.1145/3489517.3530400},
  bibsource    = {dblp computer science bibliography, https://dblp.org}
}

@inproceedings{DBLP:conf/dac/0002LG0P022,
  author       = {Hanrui Wang and
                  Zirui Li and
                  Jiaqi Gu and
                  Yongshan Ding and
                  David Z. Pan and
                  Song Han},
  editor       = {Rob Oshana},
  title        = {{QOC:} quantum on-chip training with parameter shift and gradient
                  pruning},
  booktitle    = {{DAC} '22: 59th {ACM/IEEE} Design Automation Conference, San Francisco,
                  California, USA, July 10 - 14, 2022},
  pages        = {655--660},
  publisher    = {{ACM}},
  year         = {2022},
  url          = {https://doi.org/10.1145/3489517.3530495},
  doi          = {10.1145/3489517.3530495},
  bibsource    = {dblp computer science bibliography, https://dblp.org}
}

@inproceedings{DBLP:conf/dac/LiangL0C0LRSLY024,
  author       = {Zhiding Liang and
                  Gang Liu and
                  Zheyuan Liu and
                  Jinglei Cheng and
                  Tianyi Hao and
                  Kecheng Liu and
                  Hang Ren and
                  Zhixin Song and
                  Ji Liu and
                  Fanny Ye and
                  Yiyu Shi},
  editor       = {Vivek De},
  title        = {Invited: Graph Learning for Parameter Prediction of Quantum Approximate
                  Optimization Algorithm},
  booktitle    = {Proceedings of the 61st {ACM/IEEE} Design Automation Conference, {DAC}
                  2024, San Francisco, CA, USA, June 23-27, 2024},
  pages        = {361:1--361:4},
  publisher    = {{ACM}},
  year         = {2024},
  url          = {https://doi.org/10.1145/3649329.3663523},
  doi          = {10.1145/3649329.3663523},
  bibsource    = {dblp computer science bibliography, https://dblp.org}
}

@inproceedings{DBLP:conf/dac/LiangSCHLWQWHQS23,
  author       = {Zhiding Liang and
                  Zhixin Song and
                  Jinglei Cheng and
                  Zichang He and
                  Ji Liu and
                  Hanrui Wang and
                  Ruiyang Qin and
                  Yiru Wang and
                  Song Han and
                  Xuehai Qian and
                  Yiyu Shi},
  title        = {Hybrid Gate-Pulse Model for Variational Quantum Algorithms},
  booktitle    = {60th {ACM/IEEE} Design Automation Conference, {DAC} 2023, San Francisco,
                  CA, USA, July 9-13, 2023},
  pages        = {1--6},
  publisher    = {{IEEE}},
  year         = {2023},
  url          = {https://doi.org/10.1109/DAC56929.2023.10247923},
  doi          = {10.1109/DAC56929.2023.10247923},
  bibsource    = {dblp computer science bibliography, https://dblp.org}
}

@inproceedings{DBLP:conf/dac/NiuHIJY24,
  author       = {Siyuan Niu and
                  Akel Hashim and
                  Costin Iancu and
                  Wibe Albert de Jong and
                  Ed Younis},
  editor       = {Vivek De},
  title        = {Effective Quantum Resource Optimization via Circuit Resizing in BQSKit},
  booktitle    = {Proceedings of the 61st {ACM/IEEE} Design Automation Conference, {DAC}
                  2024, San Francisco, CA, USA, June 23-27, 2024},
  pages        = {321:1--321:6},
  publisher    = {{ACM}},
  year         = {2024},
  url          = {https://doi.org/10.1145/3649329.3656534},
  doi          = {10.1145/3649329.3656534},
  bibsource    = {dblp computer science bibliography, https://dblp.org}
}

@inproceedings{DBLP:conf/focs/BoulandFLL21,
  author       = {Adam Bouland and
                  Bill Fefferman and
                  Zeph Landau and
                  Yunchao Liu},
  title        = {Noise and the Frontier of Quantum Supremacy},
  booktitle    = {62nd {IEEE} Annual Symposium on Foundations of Computer Science, {FOCS}
                  2021, Denver, CO, USA, February 7-10, 2022},
  pages        = {1308--1317},
  publisher    = {{IEEE}},
  year         = {2021},
  url          = {https://doi.org/10.1109/FOCS52979.2021.00127},
  doi          = {10.1109/FOCS52979.2021.00127},
  bibsource    = {dblp computer science bibliography, https://dblp.org}
}

@inproceedings{DBLP:conf/hpca/WangDGLPCH22,
  author       = {Hanrui Wang and
                  Yongshan Ding and
                  Jiaqi Gu and
                  Yujun Lin and
                  David Z. Pan and
                  Frederic T. Chong and
                  Song Han},
  title        = {QuantumNAS: Noise-Adaptive Search for Robust Quantum Circuits},
  booktitle    = {{IEEE} International Symposium on High-Performance Computer Architecture,
                  {HPCA} 2022, Seoul, South Korea, April 2-6, 2022},
  pages        = {692--708},
  publisher    = {{IEEE}},
  year         = {2022},
  url          = {https://doi.org/10.1109/HPCA53966.2022.00057},
  doi          = {10.1109/HPCA53966.2022.00057},
  bibsource    = {dblp computer science bibliography, https://dblp.org}
}

@inproceedings{DBLP:conf/iccad/DiBritaLHLP24,
  author       = {Nicholas S. DiBrita and
                  Daniel Leeds and
                  Yuqian Huo and
                  Jason Ludmir and
                  Tirthak Patel},
  editor       = {Jinjun Xiong and
                  Robert Wille},
  title        = {ReCon: Reconfiguring Analog Rydberg Atom Quantum Computers for Quantum
                  Generative Adversarial Networks},
  booktitle    = {Proceedings of the 43rd {IEEE/ACM} International Conference on Computer-Aided
                  Design, {ICCAD} 2024, Newark Liberty International Airport Marriott,
                  NJ, USA, October 27-31, 2024},
  pages        = {32:1--32:9},
  publisher    = {{ACM}},
  year         = {2024},
  url          = {https://doi.org/10.1145/3676536.3676697},
  doi          = {10.1145/3676536.3676697},
  bibsource    = {dblp computer science bibliography, https://dblp.org}
}

@inproceedings{DBLP:conf/iccad/LiuJSWG24,
  author       = {Yiwen Liu and
                  Qingyue Jiao and
                  Yiyu Shi and
                  Ke Wan and
                  Shangjie Guo},
  editor       = {Jinjun Xiong and
                  Robert Wille},
  title        = {A comparison on constrain encoding methods for quantum approximate
                  optimization algorithm},
  booktitle    = {Proceedings of the 43rd {IEEE/ACM} International Conference on Computer-Aided
                  Design, {ICCAD} 2024, Newark Liberty International Airport Marriott,
                  NJ, USA, October 27-31, 2024},
  pages        = {64:1--64:7},
  publisher    = {{ACM}},
  year         = {2024},
  url          = {https://doi.org/10.1145/3676536.3697126},
  doi          = {10.1145/3676536.3697126},
  bibsource    = {dblp computer science bibliography, https://dblp.org}
}

@inproceedings{DBLP:conf/pldi/XuLPLPHMPAAJ22,
  author       = {Mingkuan Xu and
                  Zikun Li and
                  Oded Padon and
                  Sina Lin and
                  Jessica Pointing and
                  Auguste Hirth and
                  Henry Ma and
                  Jens Palsberg and
                  Alex Aiken and
                  Umut A. Acar and
                  Zhihao Jia},
  editor       = {Ranjit Jhala and
                  Isil Dillig},
  title        = {Quartz: superoptimization of Quantum circuits},
  booktitle    = {{PLDI} '22: 43rd {ACM} {SIGPLAN} International Conference on Programming
                  Language Design and Implementation, San Diego, CA, USA, June 13 -
                  17, 2022},
  pages        = {625--640},
  publisher    = {{ACM}},
  year         = {2022},
  url          = {https://doi.org/10.1145/3519939.3523433},
  doi          = {10.1145/3519939.3523433},
  bibsource    = {dblp computer science bibliography, https://dblp.org}
}

@inproceedings{DBLP:conf/pldi/YuP21,
  author       = {Nengkun Yu and
                  Jens Palsberg},
  editor       = {Stephen N. Freund and
                  Eran Yahav},
  title        = {Quantum abstract interpretation},
  booktitle    = {{PLDI} '21: 42nd {ACM} {SIGPLAN} International Conference on Programming
                  Language Design and Implementation, Virtual Event, Canada, June 20-25,
                  2021},
  pages        = {542--558},
  publisher    = {{ACM}},
  year         = {2021},
  url          = {https://doi.org/10.1145/3453483.3454061},
  doi          = {10.1145/3453483.3454061},
  bibsource    = {dblp computer science bibliography, https://dblp.org}
}

@article{gilyen2018quantum,
  title={Quantum-inspired low-rank stochastic regression with logarithmic dependence on the dimension},
  author={Gily{\'e}n, Andr{\'a}s and Lloyd, Seth and Tang, Ewin},
  journal={arXiv preprint arXiv:1811.04909},
  year={2018}
}

@article{kothari2025no,
  title={No exponential quantum speedup for {$\mathrm{SIS}^\infty$} anymore},
  author={Kothari, Robin and O'Donnell, Ryan and Wu, Kewen},
  journal={arXiv preprint arXiv:2510.07515},
  year={2025}
}

@inproceedings{tang2019quantum,
  title={A quantum-inspired classical algorithm for recommendation systems},
  author={Tang, Ewin},
  booktitle={Proceedings of the 51st annual ACM SIGACT symposium on theory of computing},
  pages={217--228},
  year={2019}
}

@inproceedings{DBLP:conf/stoc/ChiaGLLTW20,
  author       = {Nai{-}Hui Chia and
                  Andr{\'{a}}s Gily{\'{e}}n and
                  Tongyang Li and
                  Han{-}Hsuan Lin and
                  Ewin Tang and
                  Chunhao Wang},
  editor       = {Konstantin Makarychev and
                  Yury Makarychev and
                  Madhur Tulsiani and
                  Gautam Kamath and
                  Julia Chuzhoy},
  title        = {Sampling-based sublinear low-rank matrix arithmetic framework for
                  dequantizing quantum machine learning},
  booktitle    = {Proceedings of the 52nd Annual {ACM} {SIGACT} Symposium on Theory
                  of Computing, {STOC} 2020, Chicago, IL, USA, June 22-26, 2020},
  pages        = {387--400},
  publisher    = {{ACM}},
  year         = {2020},
  url          = {https://doi.org/10.1145/3357713.3384314},
  doi          = {10.1145/3357713.3384314},
  bibsource    = {dblp computer science bibliography, https://dblp.org}
}

@article{jordan2025optimization,
  title={Optimization by decoded quantum interferometry},
  author={Jordan, Stephen P and Shutty, Noah and Wootters, Mary and Zalcman, Adam and Schmidhuber, Alexander and King, Robbie and Isakov, Sergei V and Khattar, Tanuj and Babbush, Ryan},
  journal={Nature},
  volume={646},
  number={8086},
  pages={831--836},
  year={2025},
  publisher={Nature Publishing Group UK London}
}

@article{schmidhuber2025quartic,
  title={Quartic quantum speedups for planted inference},
  author={Schmidhuber, Alexander and O’Donnell, Ryan and Kothari, Robin and Babbush, Ryan},
  journal={Physical Review X},
  volume={15},
  number={2},
  pages={021077},
  year={2025},
  publisher={APS}
}

@book{mezard2009information,
  title={Information, physics, and computation},
  author={Mezard, Marc and Montanari, Andrea},
  year={2009},
  publisher={Oxford University Press}
}

@inproceedings{gharibian2022dequantizing,
  title={Dequantizing the quantum singular value transformation: hardness and applications to quantum chemistry and the quantum {PCP} conjecture},
  author={Gharibian, Sevag and Le Gall, Fran{\c{c}}ois},
  booktitle={Proceedings of the 54th annual ACM SIGACT symposium on theory of computing},
  pages={19--32},
  year={2022}
}

@article{bostanci2023unitary,
  title={Unitary complexity and the {U}hlmann transformation problem},
  author={Bostanci, John and Efron, Yuval and Metger, Tony and Poremba, Alexander and Qian, Luowen and Yuen, Henry},
  journal={arXiv preprint arXiv:2306.13073},
  year={2023}
}

@inproceedings{cade2023improved,
  title={Improved {H}ardness {R}esults for the {G}uided {L}ocal {H}amiltonian {P}roblem},
  author={Cade, Chris and Folkertsma, Marten and Gharibian, Sevag and Hayakawa, Ryu and Le Gall, Fran{\c{c}}ois and Morimae, Tomoyuki and Weggemans, Jordi},
  booktitle={50th International Colloquium on Automata, Languages, and Programming (ICALP 2023)},
  pages={32--1},
  year={2023},
  organization={Schloss Dagstuhl--Leibniz-Zentrum f{\"u}r Informatik}
}

@inproceedings{aaronson2010bqp,
  title={{BQP} and the polynomial hierarchy},
  author={Aaronson, Scott},
  booktitle={Proceedings of the Forty-Second ACM Symposium on Theory of Computing},
  pages={141--150},
  year={2010}
}

@article{raz2022oracle,
  title={Oracle separation of {BQP} and {PH}},
  author={Raz, Ran and Tal, Avishay},
  journal={ACM Journal of the ACM (JACM)},
  volume={69},
  number={4},
  pages={1--21},
  year={2022},
  publisher={ACM New York, NY}
}

@article{bernstein1997quantum,
  title={Quantum {C}omplexity {T}heory},
  author={Bernstein, Ethan and Vazirani, Umesh},
  journal={{SIAM} Journal on Computing},
  volume={26},
  number={5},
  pages={1411--1473},
  year={1997},
  publisher={SIAM}
}

@inproceedings{DBLP:conf/stoc/KalaiLV023,
  author       = {Yael Kalai and
                  Alex Lombardi and
                  Vinod Vaikuntanathan and
                  Lisa Yang},
  editor       = {Barna Saha and
                  Rocco A. Servedio},
  title        = {Quantum Advantage from Any Non-local Game},
  booktitle    = {Proceedings of the 55th Annual {ACM} Symposium on Theory of Computing,
                  {STOC} 2023, Orlando, FL, USA, June 20-23, 2023},
  pages        = {1617--1628},
  publisher    = {{ACM}},
  year         = {2023},
  url          = {https://doi.org/10.1145/3564246.3585164},
  doi          = {10.1145/3564246.3585164},
  bibsource    = {dblp computer science bibliography, https://dblp.org}
}

@inproceedings{DBLP:conf/stoc/KhuranaT24,
  author       = {Dakshita Khurana and
                  Kabir Tomer},
  editor       = {Bojan Mohar and
                  Igor Shinkar and
                  Ryan O'Donnell},
  title        = {Commitments from Quantum One-Wayness},
  booktitle    = {Proceedings of the 56th Annual {ACM} Symposium on Theory of Computing,
                  {STOC} 2024, Vancouver, BC, Canada, June 24-28, 2024},
  pages        = {968--978},
  publisher    = {{ACM}},
  year         = {2024},
  url          = {https://doi.org/10.1145/3618260.3649654},
  doi          = {10.1145/3618260.3649654},
  bibsource    = {dblp computer science bibliography, https://dblp.org}
}

@inproceedings{DBLP:conf/stoc/KhuranaT25,
  author       = {Dakshita Khurana and
                  Kabir Tomer},
  editor       = {Michal Kouck{\'{y}} and
                  Nikhil Bansal},
  title        = {Founding Quantum Cryptography on Quantum Advantage, or, Towards Cryptography
                  from {\#}P Hardness},
  booktitle    = {Proceedings of the 57th Annual {ACM} Symposium on Theory of Computing,
                  {STOC} 2025, Prague, Czechia, June 23-27, 2025},
  pages        = {178--188},
  publisher    = {{ACM}},
  year         = {2025},
  url          = {https://doi.org/10.1145/3717823.3718145},
  doi          = {10.1145/3717823.3718145},
  bibsource    = {dblp computer science bibliography, https://dblp.org}
}

@inproceedings{DBLP:conf/stoc/LombardiMW24,
  author       = {Alex Lombardi and
                  Fermi Ma and
                  John Wright},
  editor       = {Bojan Mohar and
                  Igor Shinkar and
                  Ryan O'Donnell},
  title        = {A One-Query Lower Bound for Unitary Synthesis and Breaking Quantum
                  Cryptography},
  booktitle    = {Proceedings of the 56th Annual {ACM} Symposium on Theory of Computing,
                  {STOC} 2024, Vancouver, BC, Canada, June 24-28, 2024},
  pages        = {979--990},
  publisher    = {{ACM}},
  year         = {2024},
  url          = {https://doi.org/10.1145/3618260.3649650},
  doi          = {10.1145/3618260.3649650},
  bibsource    = {dblp computer science bibliography, https://dblp.org}
}

@inproceedings{DBLP:conf/vldb/GruenwaldWCGG23,
  author       = {Le Gruenwald and
                  Tobias Winker and
                  Umut {\c{C}}alikyilmaz and
                  Jinghua Groppe and
                  Sven Groppe},
  editor       = {Rajesh Bordawekar and
                  Cinzia Cappiello and
                  Vasilis Efthymiou and
                  Lisa Ehrlinger and
                  Vijay Gadepally and
                  Sainyam Galhotra and
                  Sandra Geisler and
                  Sven Groppe and
                  Le Gruenwald and
                  Alon Y. Halevy and
                  Hazar Harmouch and
                  Oktie Hassanzadeh and
                  Ihab F. Ilyas and
                  Ernesto Jim{\'{e}}nez{-}Ruiz and
                  Sanjay Krishnan and
                  Tirthankar Lahiri and
                  Guoliang Li and
                  Jiaheng Lu and
                  Wolfgang Mauerer and
                  Umar Farooq Minhas and
                  Felix Naumann and
                  M. Tamer {\"{O}}zsu and
                  El Kindi Rezig and
                  Kavitha Srinivas and
                  Michael Stonebraker and
                  Satyanarayana R. Valluri and
                  Maria{-}Esther Vidal and
                  Haixun Wang and
                  Jiannan Wang and
                  Yingjun Wu and
                  Xun Xue and
                  Mohamed Za{\"{\i}}t and
                  Kai Zeng},
  title        = {Index Tuning with Machine Learning on Quantum Computers for Large-Scale
                  Database Applications},
  booktitle    = {Joint Proceedings of Workshops at the 49th International Conference
                  on Very Large Data Bases {(VLDB} 2023), Vancouver, Canada, August
                  28 - September 1, 2023},
  series       = {{CEUR} Workshop Proceedings},
  volume       = {3462},
  publisher    = {CEUR-WS.org},
  year         = {2023},
  url          = {https://ceur-ws.org/Vol-3462/QDSM5.pdf},
  bibsource    = {dblp computer science bibliography, https://dblp.org}
}

@article{cong2019qcnn,
  title={Quantum convolutional neural networks},
  author={Cong, Iris and Choi, Soonwon and Lukin, Mikhail D},
  journal={Nature Physics},
  volume={15},
  number={12},
  pages={1273--1278},
  year={2019},
  publisher={Nature Publishing Group UK London},
doi={10.1038/s41567-019-0648-8}
}

@ARTICLE{lin2022dsqao,
  author={Lin, Wan-Hsuan and Tan, Bochen and Niu, Murphy Yuezhen and Kimko, Jason and Cong, Jason},
  journal={IEEE Journal on Emerging and Selected Topics in Circuits and Systems}, 
  title={Domain-Specific Quantum Architecture Optimization}, 
  year={2022},
  volume={12},
  number={3},
  pages={624-637},
  doi={10.1109/JETCAS.2022.3202870}}

@article{qaoa,
  title={A quantum approximate optimization algorithm},
  author={Farhi, Edward and Goldstone, Jeffrey and Gutmann, Sam},
  journal={arXiv preprint arXiv:1411.4028},
  year={2014}
}

@article{queko,
  title={Optimality study of existing quantum computing layout synthesis tools},
  author={Tan, Bochen and Cong, Jason},
  journal={IEEE Transactions on Computers},
  volume={70},
  number={9},
  pages={1363--1373},
  year={2020},
  publisher={IEEE}
}

@article{qubikos,
  title={Assessing Quantum Layout Synthesis Tools via Known Optimal-SWAP Cost Benchmarks},
  author={Ping, Shuohao and Lin, Wan-Hsuan and Tan, Daniel Bochen and Cong, Jason},
  journal={arXiv preprint arXiv:2502.08839},
  year={2025}
}

@inproceedings{zac,
  title={Reuse-aware compilation for zoned quantum architectures based on neutral atoms},
  author={Lin, Wan-Hsuan and Tan, Daniel Bochen and Cong, Jason},
  booktitle={2025 IEEE International Symposium on High Performance Computer Architecture (HPCA)},
  pages={127--142},
  year={2025},
  organization={IEEE}
}

@misc{BoulandFefferman2,
      title={Exponential improvements to the average-case hardness of BosonSampling}, 
      author={Adam Bouland and Ishaun Datta and Bill Fefferman and Felipe Hernandez},
      year={2025},
      eprint={2411.04566},
      archivePrefix={arXiv},
      primaryClass={quant-ph},
      url={https://arxiv.org/abs/2411.04566}, 
}

@misc{fefferman-linearmodes,
      title={Complexity-theoretic foundations of BosonSampling with a linear number of modes}, 
      author={Adam Bouland and Daniel Brod and Ishaun Datta and Bill Fefferman and Daniel Grier and Felipe Hernandez and Michal Oszmaniec},
      year={2025},
      eprint={2312.00286},
      archivePrefix={arXiv},
      primaryClass={quant-ph},
      url={https://arxiv.org/abs/2312.00286}, 
}

@misc{Fefferman-nonunital,
      title={Effect of non-unital noise on random circuit sampling}, 
      author={Bill Fefferman and Soumik Ghosh and Michael Gullans and Kohdai Kuroiwa and Kunal Sharma},
      year={2023},
      eprint={2306.16659},
      archivePrefix={arXiv},
      primaryClass={quant-ph},
      url={https://arxiv.org/abs/2306.16659}, 
}

@article{Fefferman-boson-algorithm,
   title={Classical algorithm for simulating experimental Gaussian boson sampling},
   volume={20},
   ISSN={1745-2481},
   url={http://dx.doi.org/10.1038/s41567-024-02535-8},
   DOI={10.1038/s41567-024-02535-8},
   number={9},
   journal={Nature Physics},
   publisher={Springer Science and Business Media LLC},
   author={Oh, Changhun and Liu, Minzhao and Alexeev, Yuri and Fefferman, Bill and Jiang, Liang},
   year={2024},
   month=jun, pages={1461–1468} }

@article{Fefferman-quantuminspired,
   title={Quantum-Inspired Classical Algorithm for Graph Problems by Gaussian Boson Sampling},
   volume={5},
   ISSN={2691-3399},
   url={http://dx.doi.org/10.1103/PRXQuantum.5.020341},
   DOI={10.1103/prxquantum.5.020341},
   number={2},
   journal={PRX Quantum},
   publisher={American Physical Society (APS)},
   author={Oh, Changhun and Fefferman, Bill and Jiang, Liang and Quesada, Nicolás},
   year={2024},
   month=may }

@article{BFNV19,
	author = {Bouland, Adam and Fefferman, Bill and Nirkhe, Chinmay and Vazirani, Umesh},
	date = {2019/02/01},
	doi = {10.1038/s41567-018-0318-2},
	id = {Bouland2019},
	isbn = {1745-2481},
	journal = {Nature Physics},
	number = {2},
	pages = {159--163},
	title = {On the complexity and verification of quantum random circuit sampling},
	url = {https://doi.org/10.1038/s41567-018-0318-2},
	volume = {15},
	year = {2019}}

@inproceedings{Dorit-RCS-algorithm,
  author       = {Dorit Aharonov and
                  Xun Gao and
                  Zeph Landau and
                  Yunchao Liu and
                  Umesh V. Vazirani},
  editor       = {Barna Saha and
                  Rocco A. Servedio},
  title        = {A Polynomial-Time Classical Algorithm for Noisy Random Circuit Sampling},
  booktitle    = {Proceedings of the 55th Annual {ACM} Symposium on Theory of Computing,
                  {STOC} 2023, Orlando, FL, USA, June 20-23, 2023},
  pages        = {945--957},
  publisher    = {{ACM}},
  year         = {2023},
  url          = {https://doi.org/10.1145/3564246.3585234},
  doi          = {10.1145/3564246.3585234},
  bibsource    = {dblp computer science bibliography, https://dblp.org}
}

@misc{schuster2024polynomialtimeclassicalalgorithmnoisy,
      title={A polynomial-time classical algorithm for noisy quantum circuits}, 
      author={Thomas Schuster and Chao Yin and Xun Gao and Norman Y. Yao},
      year={2024},
      eprint={2407.12768},
      archivePrefix={arXiv},
      primaryClass={quant-ph},
      url={https://arxiv.org/abs/2407.12768}, 
}

@article{Childs-walk,
  title = {Universal Computation by Quantum Walk},
  author = {Childs, Andrew M.},
  journal = {Phys. Rev. Lett.},
  volume = {102},
  issue = {18},
  pages = {180501},
  numpages = {4},
  year = {2009},
  month = {May},
  publisher = {American Physical Society},
  doi = {10.1103/PhysRevLett.102.180501},
  url = {https://link.aps.org/doi/10.1103/PhysRevLett.102.180501}
}

@misc{Farhi-adiabatic,
      title={Quantum Computation by Adiabatic Evolution}, 
      author={Edward Farhi and Jeffrey Goldstone and Sam Gutmann and Michael Sipser},
      year={2000},
      eprint={quant-ph/0001106},
      archivePrefix={arXiv},
      primaryClass={quant-ph},
      url={https://arxiv.org/abs/quant-ph/0001106}, 
}

@misc{aaronson-linearoptical,
      title={The Computational Complexity of Linear Optics}, 
      author={Scott Aaronson and Alex Arkhipov},
      year={2010},
      eprint={1011.3245},
      archivePrefix={arXiv},
      primaryClass={quant-ph},
      url={https://arxiv.org/abs/1011.3245}, 
}

@article{Bravyi-Gosset-Konig,
   title={Quantum advantage with shallow circuits},
   volume={362},
   ISSN={1095-9203},
   url={http://dx.doi.org/10.1126/science.aar3106},
   DOI={10.1126/science.aar3106},
   number={6412},
   journal={Science},
   publisher={American Association for the Advancement of Science (AAAS)},
   author={Bravyi, Sergey and Gosset, David and König, Robert},
   year={2018},
   month=oct, pages={308–311} }

@article{google2023suppressing,
author = "Google Quantum AI",
  title={Suppressing quantum errors by scaling a surface code logical qubit},
  journal={Nature},
  volume={614},
  number={7949},
  pages={676--681},
  year={2023},
  publisher={Nature Publishing Group UK London}
}

@article{bluvstein2024logical,
  title={Logical quantum processor based on reconfigurable atom arrays},
  author={Bluvstein, Dolev and Evered, Simon J and Geim, Alexandra A and Li, Sophie H and Zhou, Hengyun and Manovitz, Tom and Ebadi, Sepehr and Cain, Madelyn and Kalinowski, Marcin and Hangleiter, Dominik and others},
  journal={Nature},
  volume={626},
  number={7997},
  pages={58--65},
  year={2024},
  publisher={Nature Publishing Group UK London}
}

@article{google2025quantum,
author = "Google Quantum AI and Collaborators",
  title={Quantum error correction below the surface code threshold},
  journal={Nature},
  volume={638},
  number={8052},
  pages={920--926},
  year={2025},
  publisher={Nature Publishing Group UK London}
}

@inproceedings{smith2022scaling,
  title={Scaling superconducting quantum computers with chiplet architectures},
  author={Smith, Kaitlin N and Ravi, Gokul Subramanian and Baker, Jonathan M and Chong, Frederic T},
  booktitle={2022 55th IEEE/ACM International Symposium on Microarchitecture (MICRO)},
  pages={1092--1109},
  year={2022},
  organization={IEEE}
}

@article{laracuente2025modeling,
  title={Modeling short-range microwave networks to scale superconducting quantum computation},
  author={LaRacuente, Nicholas and Smith, Kaitlin N and Imany, Poolad and Silverman, Kevin L and Chong, Frederic T},
  journal={Quantum},
  volume={9},
  pages={1581},
  year={2025},
  publisher={Verein zur F{\"o}rderung des Open Access Publizierens in den Quantenwissenschaften}
}

@inproceedings{lin2024codesign,
  title={Codesign of quantum error-correcting codes and modular chiplets in the presence of defects},
  author={Lin, Sophia Fuhui and Viszlai, Joshua and Smith, Kaitlin N and Ravi, Gokul Subramanian and Yuan, Charles and Chong, Frederic T and Brown, Benjamin J},
  booktitle={Proceedings of the 29th ACM International Conference on Architectural Support for Programming Languages and Operating Systems, Volume 2},
  pages={216--231},
  year={2024}
}

@article{palmer2025boundaries,
  title={Boundaries of Acceptable Defectiveness: Redefining Surface Code Robustness under Heterogeneous Noise},
  author={Palmer, Jacob S and Smith, Kaitlin N},
  journal={arXiv preprint arXiv:2510.22001},
  year={2025}
}

@inproceedings{lin2022let,
  title={Let each quantum bit choose its basis gates},
  author={Lin, Sophia Fuhui and Sussman, Sara and Duckering, Casey and Mundada, Pranav S and Baker, Jonathan M and Kumar, Rohan S and Houck, Andrew A and Chong, Frederic T},
  booktitle={2022 55th IEEE/ACM International Symposium on Microarchitecture (MICRO)},
  pages={1042--1058},
  year={2022},
  organization={IEEE}
}

@article{jeng2025modular,
  title={Modular Compilation for Quantum Chiplet Architectures},
  author={Jeng, Mingyoung Jessica and Maruszewski, Nikola Vuk and Selna, Connor and Gavrincea, Michael and Smith, Kaitlin N and Hardavellas, Nikos},
  journal={arXiv preprint arXiv:2501.08478},
  year={2025}
}

@inproceedings{dibrita2025resq,
  title={{ResQ: A Novel Framework to Implement Residual Neural Networks on Analog Rydberg Atom Quantum Computers}},
  author={DiBrita, Nicholas S and Han, Jason and Patel, Tirthak},
  booktitle={Proceedings of the IEEE/CVF International Conference on Computer Vision},
  year={2025}
}

@inproceedings{huo2026anchor,
  title={Anchor: Reducing Temporal and Spatial Output Performance Variability on Quantum Computers},
  author={Huo, Yuqian and Leeds, Daniel and Ludmir, Jason and DiBrita, Nicholas S. and Patel, Tirthak},
  booktitle={ACM SIGMETRICS International Conference on Measurement and Modeling of Computer Systems},
  year={2026}
}

@inproceedings{ludmir2024parallax,
  title={PARALLAX: A Compiler for Neutral Atom Quantum Computers under Hardware Constraints},
  author={Ludmir, Jason and Patel, Tirthak},
  booktitle={SC24: International Conference for High Performance Computing, Networking, Storage and Analysis},
  pages={1--17},
  year={2024},
  organization={IEEE}
}

@inproceedings{huo2026nest,
  title={Three Birds with One Stone: Improving Performance, Convergence, and System Throughput with NEST},
  author={Huo, Yuqian and Quiroga, David and Kyrillidis, Anastasios and Patel, Tirthak},
  booktitle={ACM SIGMETRICS International Conference on Measurement and Modeling of Computer Systems},
  year={2026}
}

@inproceedings{ludmir2025modeling,
  title={Modeling and Simulating Rydberg Atom Quantum Computers for Hardware-Software Co-design with PachinQo},
  author={Ludmir, Jason and Huo, Yuqian and DiBrita, Nicholas S and Patel, Tirthak},
  booktitle={ACM SIGMETRICS International Conference on Measurement and Modeling of Computer Systems},
  pages={136--138},
  year={2025}
}

@misc{yuan2025cobble,
  author        = {Charles Yuan},
  archiveprefix = {arXiv},
  doi           = {10.48550/arXiv.2511.01736},
  eprint        = {2511.01736},
  primaryclass  = {cs.PL},
  title         = {{Cobble: Compiling Block Encodings for Quantum Computational Linear Algebra}},
  year          = {2025},
  note          = {arXiv preprint arXiv:2511.01736},
}

@inproceedings{yuan2024qcm,
  author    = {Charles Yuan and Agnes Villanyi and Michael Carbin},
  booktitle = {ACM SIGPLAN Conference on Object-Oriented Programming, Systems, Languages, and Applications},
  doi       = {10.1145/3649811},
  pages     = {1--28},
  title     = {{Quantum Control Machine: The Limits of Control Flow in Quantum Programming}},
  year      = {2024}
}

@inproceedings{yuan2022tower,
  author    = {Yuan, Charles and Carbin, Michael},
  booktitle = {ACM SIGPLAN Conference on Object-Oriented Programming, Systems, Languages, and Applications},
  doi       = {10.1145/3563297},
  pages     = {259--288},
  title     = {{Tower: Data Structures in Quantum Superposition}},
  year      = {2022}
}

@inproceedings{yuan2022twist,
  author    = {Charles Yuan and Christopher McNally and Michael Carbin},
  booktitle = {ACM SIGPLAN Symposium on Principles of Programming Languages},
  doi       = {10.1145/3498691},
  pages     = {1--32},
  title     = {{Twist: Sound Reasoning for Purity and Entanglement in Quantum Programs}},
  year      = {2022}
}

@inproceedings{yuan2024spire,
  author    = {Charles Yuan and Michael Carbin},
  booktitle = {ACM SIGPLAN Conference on Programming Language Design and Implementation},
  doi       = {10.1145/3656397},
  pages     = {492--517},
  title     = {{The \emph{T}-Complexity Costs of Error Correction for Control Flow in Quantum Computation}},
  year      = {2024}
}

@misc{harrigan2024,
  author        = {Matthew P. Harrigan and Tanuj Khattar and Charles Yuan and Anurudh Peduri and Noureldin Yosri and Fionn D. Malone and Ryan Babbush and Nicholas C. Rubin},
  archiveprefix = {arXiv},
  doi           = {10.48550/arXiv.2409.04643},
  eprint        = {2409.04643},
  primaryclass  = {quant-ph},
  title         = {{Expressing and Analyzing Quantum Algorithms with Qualtran}},
  year          = {2024},
  note          = {arXiv preprint arXiv:2409.04643}
}

@misc{wang2024coprimebivariatebicyclecodes,
      title={Coprime Bivariate Bicycle Codes and their Layouts on Cold
Atoms}, author={Ming Wang and Frank Mueller},
      year={2024},
      eprint={2408.10001},
      archivePrefix={arXiv},
      primaryClass={quant-ph},
      url={https://arxiv.org/abs/2408.10001},
}

@string{HPCA = "International Symposium on High Performance Computer Architecture"}

@inproceedings{wang26,
  title={Fully Parallelized BP Decoding for Quantum LDPC Codes Can
Outperform BP-OSD}, author={Ming Wang and Ang Li and Frank Mueller},
  year={2026},
  booktitle = HPCA,
  month = feb
}

@inproceedings{wilson22,
        title     = "Combining Hard and Soft Constraints in Quantum
Constraint-Satisfaction Systems", author    = 
        "Ellis Wilson and Frank Mueller and Scott Pakin", 
        booktitle = SC,
        year = 2022,
        month = nov,
        pages = {},
}

@string{SC       = "Supercomputing"}

@inproceedings{bugstahler24,
    author={Blake Burgstahlera and Frank Mueller and Scott Pakin},
  booktitle={2024 IEEE International Conference on Quantum Computing
and Engineering (QCE)},
  title={Synthesis of Approximate Parametric Circuits for
Variational Quantum Algorithms},
  year={2024},
}

@inproceedings{badrike23,
    title={QisDAX: An Open Source Bridge from Qiskit to Ion Trap
Quantum Devices}, author={Kaustubh Badrike and Aniket S. Dalvi and
Filip Mazurek and Marissa D'Onofrio and Jacob Whitlow and Tianyi Chen
and Samuel Phiri and Leon Riesebos and Kenneth R. Brown and Frank
Mueller}, year={2023},
    booktitle={IEEE International Conference on Quantum Computing and
Engineering (QCE)}, month = sep,
}

@Article{NiuSanial23,
  author = "Siyuan Niu and Aida Todri-Sanial",
  title = "Enabling multi-programming mechanism for quantum computing in the {NISQ} era",
  journal = "Quantum",
  volume = 7,
  year = 2023,
  pages = 925}

@InProceedings{NiuSanial22,
  author = "Siyuan Niu and Aida Todri-Sanial",  
  title = "How parallel circuit execution can be useful for {NISQ} computing?",
  booktitle = "2022 Design, Automation and Test in Europe Conference and Exhibition (DATE)",
  year = 2022,
  pages = "1065--1070"}

@Article{NiuSanialBronn24,
  author = "Siyuan Niu and Aida Todri-Sanial and Nicholas T. Bronn",
  title  = "Multi-qubit dynamical decoupling for enhanced crosstalk suppression",
  journal = "Quantum Science and Technology",
  volume = 9,
  number = 4,
  year = 2024}

@Unpublished{NiuKokcuMitraSzaszHashimKalloorJongIancuYounis24,
  author = "Siyuan Niu and Efekan Kokcu and Anupam Mitra and Aaron Szasz and Akel Hashim and Justin Kalloor and Wibe Albert de Jong and Costin Iancu and Ed Younis",
  title = "{AC/DC}: Automated compilation for dynamic circuits",
  note = "arXiv preprint arXiv:2412.07969",
  year = 2024}

@misc{chen2025informationcomputationgapsquantumlearning,
      title={Information-Computation Gaps in Quantum Learning via Low-Degree Likelihood}, 
      author={Sitan Chen and Weiyuan Gong and Jonas Haferkamp and Yihui Quek},
      year={2025},
      eprint={2505.22743},
      archivePrefix={arXiv},
      primaryClass={quant-ph},
    note = "Presented at QIP 2026",
      url={https://arxiv.org/abs/2505.22743}, 
}

@article{PRXQuantum.4.040337,
  title = {Learning to Predict Arbitrary Quantum Processes},
  author = {Huang, Hsin-Yuan and Chen, Sitan and Preskill, John},
  journal = {PRX Quantum},
  volume = {4},
  issue = {4},
  pages = {040337},
  numpages = {44},
  year = {2023},
  month = {Dec},
  publisher = {American Physical Society},
  doi = {10.1103/PRXQuantum.4.040337},
  url = {https://link.aps.org/doi/10.1103/PRXQuantum.4.040337}
}

@Unpublished{ChenCutlerHuang25,
    author = "Sitan Chen and Jordan Cotler and Hsin-Yuan Huang",
    title = "Quantum Probe Tomography",
    note = "arXiv, \url{https://arxiv.org/pdf/2510.08499}"}

@INPROCEEDINGS{10756089,
  author={Chen, Sitan and Gong, Weiyuan and Ye, Qi},
  booktitle={2024 IEEE 65th Annual Symposium on Foundations of Computer Science (FOCS)}, 
  title={Optimal Tradeoffs for Estimating Pauli Observables}, 
  year={2024},
  volume={},
  number={},
  pages={1086-1105},
    note = "Also at QIP 2025",
  doi={10.1109/FOCS61266.2024.00072}}

@inproceedings{10.1145/3717823.3718191,
author = {Chen, Sitan and Gong, Weiyuan and Ye, Qi and Zhang, Zhihan},
title = {Stabilizer Bootstrapping: A Recipe for Efficient Agnostic Tomography and Magic Estimation},
year = {2025},
isbn = {9798400715105},
publisher = {Association for Computing Machinery},
address = {New York, NY, USA},
url = {https://doi.org/10.1145/3717823.3718191},
doi = {10.1145/3717823.3718191},
booktitle = {Proceedings of the 57th Annual ACM Symposium on Theory of Computing},
pages = {429–438},
numpages = {10},
location = {Prague, Czechia},
series = {STOC '25},
note = "Also at QIP 2025",
}

@InProceedings{pmlr-v291-chen25c,
  title = 	 {Predicting quantum channels over general product distributions},
  author =       {Chen, Sitan and {de Dios Pont}, Jaume and Hsieh, Jun-Ting and Huang, Hsin-Yuan and Lange, Jane and Li, Jerry},
  booktitle = 	 {Proceedings of Thirty Eighth Conference on Learning Theory},
  pages = 	 {986--1007},
  year = 	 {2025},
  editor = 	 {Haghtalab, Nika and Moitra, Ankur},
  volume = 	 {291},
  series = 	 {Proceedings of Machine Learning Research},
  month = 	 {30 Jun--04 Jul},
  publisher =    {PMLR},
  url = 	 {https://proceedings.mlr.press/v291/chen25c.html}
}

@misc{leng2025subexponentialquantumspeedupoptimization,
      title={({Sub})Exponential Quantum Speedup for Optimization}, 
      author={Jiaqi Leng and Kewen Wu and Xiaodi Wu and Yufan Zheng},
      year={2025},
      eprint={2504.14841},
      archivePrefix={arXiv},
      primaryClass={quant-ph},
      url={https://arxiv.org/abs/2504.14841}, 
}

@article{Leng2025expandinghardware,
  doi = {10.22331/q-2025-09-11-1857},
  url = {https://doi.org/10.22331/q-2025-09-11-1857},
  title = {Expanding {H}ardware-{E}fficiently {M}anipulable {H}ilbert {S}pace via {H}amiltonian {E}mbedding},
  author = {Leng, Jiaqi and Li, Joseph and Peng, Yuxiang and Wu, Xiaodi},
  journal = {{Quantum}},
  issn = {2521-327X},
  publisher = {{Verein zur F{\"{o}}rderung des Open Access Publizierens in den Quantenwissenschaften}},
  volume = {9},
  pages = {1857},
  month = sep,
  year = {2025}
}

@Unpublished{Chaudhary25,
  author = "S. Chaudhary and J. Cheng and S. Kushnir and J. Leng and P. Liu and Y. Peng and H. Wang and X. Wu",
  title = "Quantum-Inspired Hamiltonian Descent for Mixed-Integer Quadratic Programming",  note = "Presented at NeurIPS Workshop on GPU-Accelerated and Scalable Optimization (ScaleOpt)",
    year = 2025}

@InProceedings{10.1007/978-3-031-07082-2_25,
author="Chung, Kai-Min
and Lee, Yi
and Lin, Han-Hsuan
and Wu, Xiaodi",
editor="Dunkelman, Orr
and Dziembowski, Stefan",
title="Constant-Round Blind Classical Verification ofÂ Quantum Sampling",
booktitle="Advances in Cryptology -- EUROCRYPT 2022",
year="2022",
publisher="Springer International Publishing",
address="Cham",
pages="707--736",
isbn="978-3-031-07082-2"
}

@article{PhysRevLett.125.150504,
  title = {Simulating Large Quantum Circuits on a Small Quantum Computer},
  author = {Peng, Tianyi and Harrow, Aram W. and Ozols, Maris and Wu, Xiaodi},
  journal = {Phys. Rev. Lett.},
  volume = {125},
  issue = {15},
  pages = {150504},
  numpages = {6},
  year = {2020},
  month = {Oct},
  publisher = {American Physical Society},
  doi = {10.1103/PhysRevLett.125.150504},
  url = {https://link.aps.org/doi/10.1103/PhysRevLett.125.150504}
}

@inproceedings{DBLP:conf/date/AlamA25,
  author={Shamiul Alam and Ahmedullah Aziz},
  title={Ferroelectric-Superconducting Synergy for Future Computing},
  year=2025,
  pages="1--6",
  url={https://doi.org/10.23919/DATE64628.2025.10993183},
  booktitle={DATE}
}

@ARTICLE{2024JAP...135a4903A,
       author = {{Alam}, Shamiul and {Hossain}, Md Shafayat and {Ni}, Kai and {Narayanan}, Vijaykrishnan and {Aziz}, Ahmedullah},
        title = "{Voltage-controlled cryogenic Boolean logic gates based on ferroelectric SQUID and heater cryotron}",
      journal = {Journal of Applied Physics},
         year = 2024,
        month = jan,
       volume = {135},
       number = {1},
          eid = {014903},
        pages = {014903},
          doi = {10.1063/5.0172531},
archivePrefix = {arXiv},
       eprint = {2212.08202},
 primaryClass = {physics.app-ph},
       adsurl = {https://ui.adsabs.harvard.edu/abs/2024JAP...135a4903A}
}

@article{10.1063/5.0170187,
    author = {Alam, Shamiul and Rampini, Dana S. and Oripov, Bakhrom G. and McCa
ughan, Adam N. and Aziz, Ahmedullah},
    title = {Cryogenic reconfigurable logic with superconducting heater cryotron
: Enhancing area efficiency and enabling camouflaged processors},
    journal = {Applied Physics Letters},
    volume = {123},
    number = {15},
    pages = {152603},
    year = {2023},
    month = {10},
    issn = {0003-6951},
    doi = {10.1063/5.0170187},
    url = {https://doi.org/10.1063/5.0170187},
    eprint = {https://pubs.aip.org/aip/apl/article-pdf/doi/10.1063/5.0170187/18163568/152603_1_5.0170187.pdf},
}

@Article{Alam23, 
  author = "Shamiul Alam and Md Shafayat Hossain and Srivatsa Rangachar Srinivasa and Ahmedullah Aziz",
  title = "Cryogenic memory technologies",
  journal = "Nat Electron",
  volume = 6,
  pages = "185--198",
  year = 2023}

@article{10.1063/5.0060716,
    author = {Alam, Shamiul and Hossain, Md Shafayat and Aziz, Ahmedullah},
    title = {A cryogenic memory array based on superconducting memristors},
    journal = {Applied Physics Letters},
    volume = {119},
    number = {8},
    pages = {082602},
    year = {2021},
    month = {08},
    issn = {0003-6951},
    doi = {10.1063/5.0060716},
    url = {https://doi.org/10.1063/5.0060716},
    eprint = {https://pubs.aip.org/aip/apl/article-pdf/doi/10.1063/5.0060716/14551138/082602_1_online.pdf},
}

@inproceedings{wilson21,
        title     = "Empirical Evaluation of  Circuit Approximations
        on Noisy Quantum Devices",
        author    = "Ellis Wilson and Frank Mueller and Lindsay
Bassman and Costin Iancu", booktitle = SC,
        year = 2021,
        month = nov,
        pages = {},
}

@inproceedings{maurya2025decoder,
  title={decoder-bench: Benchmarking Decoders for Quantum Error Correction},
  author={Maurya, Satvik and Viszlai, Joshua and Raveendran, Nithin and Das, Poulami and Tannu, Swamit},
  booktitle={2025 IEEE International Symposium on Workload Characterization (IISWC)},
  pages={286--295},
  year={2025},
  organization={IEEE}
}

@inproceedings{das2023imitation,
  title={The imitation game: Leveraging copycats for robust native gate selection in nisq programs},
  author={Das, Poulami and Kessler, Eric and Shi, Yunong},
  booktitle={2023 IEEE International Symposium on High-Performance Computer Architecture (HPCA)},
  pages={787--801},
  year={2023},
  organization={IEEE}
}

@inproceedings{wang2024qoncord,
  title={Qoncord: A Multi-Device Job Scheduling Framework for Variational Quantum Algorithms},
  author={Wang, Meng and Das, Poulami and Nair, Prashant J},
  booktitle={2024 57th IEEE/ACM International Symposium on Microarchitecture (MICRO)},
  pages={735--749},
  year={2024},
  organization={IEEE}
}

@inproceedings{das2022lilliput,
  title={Lilliput: a lightweight low-latency lookup-table decoder for near-term
+quantum error correction},
  author={Das, Poulami and Locharla, Aditya and Jones, Cody},
  booktitle={Proceedings of the 27th ACM International Conference on Architectural Support for Programming Languages and Operating Systems},
  pages={541--553},
  year={2022}
}

@misc{jiao2025mediqganquantuminspiredganhigh,
      title={{MediQ-GAN}: Quantum-Inspired {GAN} for High Resolution Medical Image Generation}, 
      author={Qingyue Jiao and Yongcan Tang and Jun Zhuang and Jason Cong and Yiyu Shi},
      year={2025},
      eprint={2506.21015},
      archivePrefix={arXiv},
      primaryClass={cs.CV},
      url={https://arxiv.org/abs/2506.21015}, 
}

@ARTICLE{10402000,
  author={Liang, Zhiding and Cheng, Jinglei and Ren, Hang and Wang, Hanrui and Hua, Fei and Song, Zhixin and Ding, Yongshan and Chong, Frederic T. and Han, Song and Qian, Xuehai and Shi, Yiyu},
  journal={IEEE Transactions on Computer-Aided Design of Integrated Circuits and Systems}, 
  title={NAPA: Intermediate-Level Variational Native-Pulse Ansatz for Variational Quantum Algorithms}, 
  year={2024},
  volume={43},
  number={6},
  pages={1834-1847},
  doi={10.1109/TCAD.2024.3355277}}

@InProceedings{Groppe22,
  author = "S. Groppe and J. Groppe and U. Calikyilmaz and T. Winker and L. Gruenwald",
  title = "Quantum Data Management and Quantum Machine Learning for Data Management: State-of-the-Art and Open Challenges",
  booktitle = "The EAI International Conference on Intelligent Systems and Machine Learning (EAI ICISML)", 
  year = 2022}

@InProceedings{Winker23,
  author = "T. Winker and U. Çalıkyılmaz and L. Gruenwald and S. Groppe",
  title = "Quantum machine learning for join order optimization using variational quantum circuits", 
  booktitle = "The International Workshop on Big Data in Emergent Distributed Environments (BiDEDE)", 
  year = 2023}

@Article{fi16120439,
AUTHOR = {Barbosa, Diogo and Gruenwald, Le and D'Orazio, Laurent and Bernardino, Jorge},
TITLE = {QRLIT: Quantum Reinforcement Learning for Database Index Tuning},
JOURNAL = {Future Internet},
VOLUME = {16},
YEAR = {2024},
NUMBER = {12},
ARTICLE-NUMBER = {439},
URL = {https://www.mdpi.com/1999-5903/16/12/439},
ISSN = {1999-5903},
DOI = {10.3390/fi16120439}
}

@misc{yu2025quantumcentricalgorithmsamplebasedkrylov,
      title={Quantum-Centric Algorithm for Sample-Based Krylov Diagonalization}, 
      author={Jeffery Yu and Javier Robledo Moreno and Joseph T. Iosue and Luke Bertels and Daniel Claudino and Bryce Fuller and Peter Groszkowski and Travis S. Humble and Petar Jurcevic and William Kirby and Thomas A. Maier and Mario Motta and Bibek Pokharel and Alireza Seif and Amir Shehata and Kevin J. Sung and Minh C. Tran and Vinay Tripathi and Antonio Mezzacapo and Kunal Sharma},
      year={2025},
      eprint={2501.09702},
      archivePrefix={arXiv},
      primaryClass={quant-ph},
      url={https://arxiv.org/abs/2501.09702}, 
}

@misc{shah2025quantumenabledbiomarkerdiscoveryoutlook,
      title={Toward Quantum-Enabled Biomarker Discovery: An Outlook from Q4Bio}, 
      author={Dhirpal Shah and Mariesa Teo and Ryan A. Robinett and Sophia Madejski and Zachary Morrell and Siddhi Ramesh and Colin Campbell and Bharath Thotakura and Victory Omole and Ben Hall and Aram W. Harrow and Teague Tomesh and Alexander T. Pearson and Frederic T. Chong and Samantha J. Riesenfeld},
      year={2025},
      eprint={2509.25904},
      archivePrefix={arXiv},
      primaryClass={quant-ph},
      url={https://arxiv.org/abs/2509.25904}, 
}

@INPROCEEDINGS{das2022afs,
  author={Das, Poulami and Pattison, Christopher A. and Manne, Srilatha and Carmean, Douglas M. and Svore, Krysta M. and Qureshi, Moinuddin and Delfosse, Nicolas},
  booktitle={HPCA}, 
  title = {AFS: Accurate, Fast, and Scalable Error-Decoding for Fault-Tolerant Quantum Computers}, 
  year = {2022},
 }

@inproceedings{vittal2023astrea,
  title={Astrea: Accurate Quantum Error-Decoding via Practical Minimum-Weight Perfect-Matching},
  author={Vittal, Suhas and Das, Poulami and Qureshi, Moinuddin},
  booktitle={ISCA-50},
  year={2023}
}

@inproceedings{eraser,
author = {Vittal, Suhas and Das, Poulami and Qureshi, Moinuddin},
title = {ERASER: Towards Adaptive Leakage Suppression for Fault-Tolerant Quantum Computing},
year = {2023},
isbn = {9798400703294},
publisher = {Association for Computing Machinery},
address = {New York, NY, USA},
url = {https://doi.org/10.1145/3613424.3614251},
doi = {10.1145/3613424.3614251},
booktitle = {Proceedings of the 56th Annual IEEE/ACM International Symposium on Microarchitecture},
pages = {509–525},
numpages = {17},
location = {Toronto, ON, Canada},
series = {MICRO '23}
}

@inproceedings{flagproxynetwork,
author = {Vittal, Suhas and Javadi-Abhari, Ali and Cross, Andrew W. and Bishop, Lev S. and Qureshi, Moinuddin},
title = {Flag-Proxy Networks: Overcoming the Architectural, Scheduling and Decoding Obstacles of Quantum LDPC Codes},
year = {2024},
publisher = {IEEE Press},
url = {https://doi.org/10.1109/MICRO61859.2024.00059},
doi = {10.1109/MICRO61859.2024.00059},
booktitle = {Proceedings of the 2024 57th IEEE/ACM International Symposium on Microarchitecture},
pages = {718–734},
numpages = {17},
location = {Austin, TX, USA},
series = {MICRO '24}
}

@inproceedings{FNM,
  title={Mitigating measurement errors in quantum computers by exploiting state-dependent bias},
  author={Tannu, Swamit S and Qureshi, Moinuddin K},
  booktitle={Proceedings of the 52nd Annual IEEE/ACM International Symposium on Microarchitecture},
  pages={279--290},
  year={2019}
}

@inproceedings{tannu2017taming,
  title={Taming the instruction bandwidth of quantum computers via hardware-managed error correction},
  author={Tannu, Swamit S and Myers, Zachary A and Nair, Prashant J and Carmean, Douglas M and Qureshi, Moinuddin K},
  booktitle={Proceedings of the 50th Annual IEEE/ACM International Symposium on Microarchitecture},
  year={2017},
  organization={IEEE}
}

@inproceedings{tannu2018case,
  title={Not all qubits are created equal: a case for variability-aware policies for NISQ-era quantum computers},
  author={Tannu, Swamit S and Qureshi, Moinuddin},
  booktitle={Proceedings of the Twenty-Fourth International Conference on Architectural Support for Programming Languages and Operating Systems},
  pages={987--999},
  year={2019},
  organization={ACM}
}

@inproceedings{micro1,
  title={Ensemble of diverse mappings: Improving reliability of quantum computers by orchestrating dissimilar mistakes},
  author={Tannu, Swamit S and Qureshi, Moinuddin},
  booktitle={Proceedings of the 52nd Annual IEEE/ACM International Symposium on Microarchitecture},
  pages={253--265},
  year={2019}
}

@inproceedings{jigsaw,
  title={Jigsaw: Boosting fidelity of nisq programs via measurement subsetting},
  author={Das, Poulami and Tannu, Swamit and Qureshi, Moinuddin},
  booktitle={MICRO-54: 54th Annual IEEE/ACM International Symposium on Microarchitecture},
  pages={937--949},
  year={2021}
}

@inproceedings{das2019case,
  title={A case for multi-programming quantum computers},
  author={Das, Poulami and Tannu, Swamit S and Nair, Prashant J and Qureshi, Moinuddin},
  booktitle={Proceedings of the 52nd Annual IEEE/ACM International Symposium on Microarchitecture},
  pages={291--303},
  year={2019}
}

@inproceedings{adapt,
  title={Adapt: Mitigating idling errors in qubits via adaptive dynamical decoupling},
  author={Das, Poulami and Tannu, Swamit and Dangwal, Siddharth and Qureshi, Moinuddin},
  booktitle={MICRO-54: 54th Annual IEEE/ACM International Symposium on Microarchitecture},
  pages={950--962},
  year={2021}
}

@inproceedings{elivagar,
    title={Élivágar: Efficient Quantum Circuit Search for Classification},
    author={Sashwat Anagolum and Narges Alavisamani and Poulami Das and Moinuddin Qureshi and Yunong Shi},
    year={2024},
 booktitle={Proceedings of the 29th ACM International Conference on Architectural Support for Programming Languages and Operating Systems (ASPLOS)}
}

@inproceedings{ayanzadeh2023frozenqubits,
  title={FrozenQubits: Boosting Fidelity of QAOA by Skipping Hotspot Nodes},
  author={Ayanzadeh, Ramin and Alavisamani, Narges and Das, Poulami and Qureshi, Moinuddin},
  booktitle={Proceedings of the 28th ACM International Conference on Architectural Support for Programming Languages and Operating Systems, Volume 2},
  pages={311--324},
  year={2023}
}

@misc{liu2025riscqgeneratorrealtimequantum,
      title={RISC-Q: A Generator for Real-Time Quantum Control System-on-Chips Compatible with RISC-V}, 
      author={Junyi Liu and Yi Lee and Haowei Deng and Connor Clayton and Gengzhi Yang and Xiaodi Wu},
      year={2025},
      eprint={2505.14902},
      archivePrefix={arXiv},
      primaryClass={cs.AR},
      url={https://arxiv.org/abs/2505.14902}, 
}

@inproceedings{10.1145/3341302.3342070,
author = {Dahlberg, Axel and Skrzypczyk, Matthew and Coopmans, Tim and Wubben, Leon and Rozpundefineddek, Filip and Pompili, Matteo and Stolk, Arian and Pawe\l{}czak, Przemys\l{}aw and Knegjens, Robert and de Oliveira Filho, Julio and Hanson, Ronald and Wehner, Stephanie},
title = {A link layer protocol for quantum networks},
year = {2019},
isbn = {9781450359566},
publisher = {Association for Computing Machinery},
address = {New York, NY, USA},
url = {https://doi.org/10.1145/3341302.3342070},
doi = {10.1145/3341302.3342070},
booktitle = {Proceedings of the ACM Special Interest Group on Data Communication},
pages = {159–173},
numpages = {15},
location = {Beijing, China},
series = {SIGCOMM '19}
}

@article{wehner2018quantum,
  title        = {Quantum Internet: A Vision for the Road Ahead},
  author       = {Wehner, Stephanie and Elkouss, David and Hanson, Ronald},
  journal      = {Science},
  volume       = {362},
  number       = {6412},
  pages        = {eaam9288},
  year         = {2018},
  doi          = {10.1126/science.aam9288},
}

@article{delle2025operating,
  title        = {An operating system for executing applications on quantum network nodes},
  author       = {Delle Donne, Carlo and Iuliano, Mariagrazia and van der Vecht, Bart and Ferreira, Guilherme Maciel and Jirovská, Hana and van der Steenhoven, Thom J. W. and Dahlberg, Axel and Skrzypczyk, Matt and Fioretto, Dario and Teller, Markus and Filippov, Pavel and Rodríguez-Pardo Montblanch, Alejandro R.-P. and Fischer, Julius and van Ommen, Benjamin H. and Demetriou, Nicolas and Leichtle, Dominik and Music, Luka and Ollivier, Harold and te Raa, Ingmar and Kozlowski, Wojciech and Taminiau, Tim H. T. and Pawełczak, Przemysław and Northup, Tracy E. and Hanson, Ronald and Wehner, Stephanie},
  journal      = {Nature},
  volume       = {639},
  pages        = {321--328},
  year         = {2025},
  doi          = {10.1038/s41586-025-08704-w},
}

@inproceedings{shi2020concurrent,
  title        = {Concurrent Entanglement Routing for Quantum Networks: Model and Designs},
  author       = {Shouqian Shi and Chen Qian},
  booktitle    = {Proceedings of the Annual Conference of the ACM Special Interest Group on Data Communication (SIGCOMM ’20)},
  year         = {2020},
  pages        = {62--75},
  doi          = {10.1145/3387514.3405853},
}

@InProceedings{li_et_al:LIPIcs.ECOOP.2024.24,
  author =	{Li, Liyi and Zhu, Mingwei and Cleaveland, Rance and Nicolellis, Alexander and Lee, Yi and Chang, Le and Wu, Xiaodi},
  title =	{{Qafny: A Quantum-Program Verifier}},
  booktitle =	{38th European Conference on Object-Oriented Programming (ECOOP 2024)},
  pages =	{24:1--24:31},
  series =	{Leibniz International Proceedings in Informatics (LIPIcs)},
  ISBN =	{978-3-95977-341-6},
  ISSN =	{1868-8969},
  year =	{2024},
  volume =	{313},
  editor =	{Aldrich, Jonathan and Salvaneschi, Guido},
  publisher =	{Schloss Dagstuhl -- Leibniz-Zentrum f{\"u}r Informatik},
  address =	{Dagstuhl, Germany},
  URL =		{https://drops.dagstuhl.de/entities/document/10.4230/LIPIcs.ECOOP.2024.24},
  URN =		{urn:nbn:de:0030-drops-208735},
  doi =		{10.4230/LIPIcs.ECOOP.2024.24}
}

@misc{clayton2024efficientroutingquantumnetworks,
      title={Efficient Routing on Quantum Networks using Adaptive Clustering}, 
      author={Connor Clayton and Xiaodi Wu and Bobby Bhattacharjee},
      howpublished={In the Proceedings of The 32nd IEEE International Conference on Network Protocols},
      year={2024},
      eprint={2410.23007},
      archivePrefix={arXiv},
      primaryClass={quant-ph},
      url={https://arxiv.org/abs/2410.23007}, 
}

@article{deng-popl-2024,
author = {Deng, Haowei and Tao, Runzhou and Peng, Yuxiang and Wu, Xiaodi},
title = {A Case for Synthesis of Recursive Quantum Unitary Programs},
year = {2024},
issue_date = {January 2024},
publisher = {Association for Computing Machinery},
address = {New York, NY, USA},
volume = {8},
number = {POPL},
url = {https://doi.org/10.1145/3632901},
doi = {10.1145/3632901},
journal = {Proc. ACM Program. Lang.},
month = jan,
articleno = {59},
numpages = {30}
}

@misc{kushnir2024qhdoptsoftwarenonlinearoptimization,
      title={{QHDOPT: A Software for Nonlinear Optimization with Quantum {Hamiltonian} Descent}}, 
      author={Samuel Kushnir and Jiaqi Leng and Yuxiang Peng and Lei Fan and Xiaodi Wu},
      year={2024},
      eprint={2409.03121},
      archivePrefix={arXiv},
      primaryClass={quant-ph},
      url={https://arxiv.org/abs/2409.03121}, 
      howpublished = {arXiv:2409.03121, to appear in INFORMS Journal on Computing. Github: https://github.com/jiaqileng/QHDOPT}
}

@misc{leng2020qhd,
    author={Jiaqi Leng and Ethan Hickman and Joseph Li and Xiaodi Wu},
    title={Quantum {Hamiltonian} Descent}, 
    howpublished={arXiv: 2303.01471. Visualization and Data available at https://jiaqileng.github.io/quantum-hamiltonian-descent/},
}

@inproceedings{Peng-PLDI22,
author = {Peng, Yuxiang and Ying, Mingsheng and Wu, Xiaodi},
title = {Algebraic Reasoning of Quantum Programs via Non-Idempotent Kleene Algebra},
year = {2022},
isbn = {9781450392655},
publisher = {Association for Computing Machinery},
address = {New York, NY, USA},
url = {https://doi.org/10.1145/3519939.3523713},
doi = {10.1145/3519939.3523713},
booktitle = {Proceedings of the 43rd ACM SIGPLAN International Conference on Programming Language Design and Implementation},
pages = {657–670},
numpages = {14},
location = {San Diego, CA, USA},
series = {PLDI 2022}
}

@misc{leng-daqc-neurips22, 
  author={Jiaqi Leng and Yuxiang Peng and Yi-Ling Qiao and Ming Lin and Xiaodi Wu},
  title={Differentiable Analog Quantum Computing for Optimization and Control},
  howpublished={Spotlight Talk at the 36th Conference on Neural Information Processing Systems (NeurIPS'22). Also available at arXiv:2210.15812.}
}

@misc{peng-simuq-2022, 
  author={Yuxiang Peng and Jacob Young and Pengyu Liu and Xiaodi Wu},
  title={{SimuQ}: A Domain-Specific Language For Quantum Simulation With Analog Compilation},
  howpublished={POPL 2024. ArXiv: 2303.02775.  Website: https://pickspeng.github.io/SimuQ/.}
}

@INPROCEEDINGS{hung19qrobust,
  AUTHOR = {Shih-Han Hung and Kesha Hietala and Shaopeng Zhu and Mingsheng Ying and Michael Hicks and Xiaodi Wu},
  TITLE = {Quantitative Robustness Analysis of Quantum Programs},
  BOOKTITLE = {Proceedings of the {ACM} Conference on Principles of Programming Languages (POPL)},
  YEAR = {2019},
  MONTH = JAN
}

@inproceedings{zhu-pldi20,
author = {Zhu, Shaopeng and Hung, Shih-Han and Chakrabarti, Shouvanik and Wu, Xiaodi},
title = {On the Principles of Differentiable Quantum Programming Languages},
year = {2020},
isbn = {9781450376136},
publisher = {Association for Computing Machinery},
address = {New York, NY, USA},
url = {https://doi.org/10.1145/3385412.3386011},
doi = {10.1145/3385412.3386011},
booktitle = {Proceedings of the 41st ACM SIGPLAN Conference on Programming Language Design and Implementation},
pages = {272–285},
numpages = {14},
location = {London, UK},
series = {PLDI 2020}
}

@inproceedings{ying-popl17,
author = {Ying, Mingsheng and Ying, Shenggang and Wu, Xiaodi},
title = {Invariants of Quantum Programs: Characterisations and Generation},
year = {2017},
isbn = {9781450346603},
publisher = {Association for Computing Machinery},
address = {New York, NY, USA},
url = {https://doi.org/10.1145/3009837.3009840},
doi = {10.1145/3009837.3009840},
booktitle = {Proceedings of the 44th ACM SIGPLAN Symposium on Principles of Programming Languages},
pages = {818–832},
numpages = {15},
location = {Paris, France},
series = {POPL '17}
}

@article{fang-unbounded-2023,
author = {Fang, Wang and Ying, Mingsheng and Wu, Xiaodi},
title = {Differentiable Quantum Programming with Unbounded Loops},
year = {2023},
publisher = {Association for Computing Machinery},
address = {New York, NY, USA},
issn = {1049-331X},
url = {https://doi.org/10.1145/3617178},
doi = {10.1145/3617178},
note = {Just Accepted},
journal = {ACM Trans. Softw. Eng. Methodol.},
month = {aug}
}

@article{peng-shor-2023,
	author = {Yuxiang Peng and Kesha Hietala and Runzhou Tao and Liyi Li and Robert Rand and Michael Hicks and Xiaodi Wu},
	journal = {Proceedings of the National Academy of Sciences},
	number = {21},
	pages = {e2218775120},
	title = {A formally certified end-to-end implementation of Shor's factorization algorithm},
	volume = {120},
	year = {2023}}

@article{li-oracle-oopsla2022,
author = {Li, Liyi and Voichick, Finn and Hietala, Kesha and Peng, Yuxiang and Wu, Xiaodi and Hicks, Michael},
title = {Verified Compilation of Quantum Oracles},
year = {2022},
issue_date = {October 2022},
publisher = {Association for Computing Machinery},
address = {New York, NY, USA},
volume = {6},
number = {OOPSLA2},
url = {https://doi.org/10.1145/3563309},
doi = {10.1145/3563309},
journal = {Proc. ACM Program. Lang.},
month = {oct},
articleno = {146},
numpages = {27}
}

\end{document}